\documentclass[iop,apj]{emulateapj_mod}

\pdfoutput=1

\usepackage{hyperref}
\usepackage{bigstrut}
\usepackage{mathptmx}

\defcitealias{homan2010}{H10}

\newcommand{\src}{\mbox{XTE J1701--462}}
\newcommand{\cyg}{\mbox{Cyg X-2}}
\newcommand{\cir}{\mbox{Cir X-1}}
\newcommand{\gx}{\mbox{GX 13+1}}
\newcommand{\maxi}{\mbox{MAXI J0556--332}}

\newcommand{\xte}{\textit{RXTE}}
\newcommand{\sw}{\textit{Swift}}
\newcommand{\cxo}{\textit{Chandra}}
\newcommand{\xmm}{\textit{XMM-Newton}}

\shorttitle{Fridriksson, Homan, \& Remillard}
\shortauthors{Fridriksson, Homan, \& Remillard}

\journalinfo{\textsc{The Astrophysical Journal}, 809:52 (21pp), 2015 August 10}
\slugcomment{Received 2014 September 12; accepted 2015 March 29; published 2015 August 10}

\begin{document}

\title{Common Patterns in the Evolution between the Luminous Neutron~Star~Low-Mass~X-ray~Binary~Subclasses}

\author{Joel~K.~Fridriksson\altaffilmark{1,2}, Jeroen~Homan\altaffilmark{2,3}, and Ronald~A.~Remillard\altaffilmark{2}}

\affil{\altaffilmark{1}Anton Pannekoek Institute for Astronomy, University of Amsterdam, Science Park 904, 1098 XH Amsterdam, The Netherlands; J.K.Fridriksson@uva.nl\\ \altaffilmark{2}Kavli Institute for Astrophysics and Space Research, Massachusetts Institute of Technology, 77 Massachusetts Avenue, Cambridge, MA 02139, USA\\\altaffilmark{3}SRON, Netherlands Institute for Space Research, Sorbonnelaan 2, 3584 CA Utrecht, The Netherlands}

\begin{abstract}

The X-ray transient \src\ was the first source observed to evolve through all known subclasses of low-magnetic-field neutron star low-mass X-ray binaries (NS-LMXBs), as a result of large changes in its mass accretion rate. To investigate to what extent similar evolution is seen in other NS-LMXBs we have performed a detailed study of the color--color and hardness--intensity diagrams (CDs and HIDs) of \cyg, \cir, and \gx---three luminous X-ray binaries, containing weakly magnetized neutron stars, known to exhibit strong secular changes in their CD/HID tracks. Using the full set of \textit{Rossi X-ray Timing Explorer} Proportional Counter Array data collected for the sources over the 16~year duration of the mission, we show that \cyg\ and \cir\ display CD/HID evolution with close similarities to \src. Although \gx\ shows behavior that is in some ways unique, it also exhibits similarities to \src, and we conclude that its overall CD/HID properties strongly indicate that it should be classified as a Z source, rather than as an atoll source. We conjecture that the secular evolution of \cyg, \cir, and \gx---illustrated by sequences of CD/HID tracks we construct---arises from changes in the mass accretion rate.  Our results strengthen previous suggestions that within single sources Cyg-like Z source behavior takes place at higher luminosities and mass accretion rates than Sco-like Z behavior, and lend support to the notion that the mass accretion rate is the primary physical parameter distinguishing the various NS-LMXB subclasses.

\end{abstract}


\section{Introduction}\label{sec:intro}

In neutron star low-mass X-ray binaries (NS-LMXBs), a neutron star accretes matter from a low-mass companion star. The vast majority of NS-LMXBs contains weakly magnetized neutron stars, and these systems are traditionally divided into two main subclasses based on their correlated spectral and timing properties: the Z sources and atoll sources \citep{hasinger1989}. The names derive from the shapes of the tracks the sources trace out in the X-ray color--color and hardness--intensity diagrams (CDs and HIDs). The Z sources are the more luminous, with near-Eddington X-ray luminositities, whereas the atoll sources are thought to in general have $L_\mathrm{X}\lesssim0.5L_\mathrm{Edd}$. Further divisions of these two main NS-LMXB types have also been made; in particular, the Z sources have been divided into the Cyg-like and Sco-like subtypes (\citealt{kuulkers1997} and references therein). 

In addition to differences in X-ray luminosity, rapid-variability characteristics, and spectral properties (see, e.g., \citealt{vanderklis2006} and references therein), the NS-LMXB subclasses also differ with respect to behavior in the radio band \citep{migliari2006} and the rates and properties of type I X-ray bursts exhibited \citep{galloway2008}. Understanding what physical factors underlie this variety in characteristics has been a long-standing problem in the study of X-ray binaries. It has long been clear that the mass accretion rate must play an important role (e.g., it seems evident that Z sources must in general accrete at a higher rate than atoll sources, given the former's significantly higher luminosity), but exactly what that role is, and where other physical parameters enter the picture, has been debated.

Another long-standing question concerns the physical nature of the motion of Z sources in the CD/HID. The Z sources trace out characteristic tracks in the CD and HID (consisting of the so-called horizontal, normal, and flaring branches) on typical timescales of hours to a few days. Initially, it was assumed that motion along the Z track was driven by changes in the mass accretion rate, with $\dot{M}$ increasing monotonically from the tip of the horizontal branch to the tip of the flaring branch (e.g., \citealt{hasinger1990}), but later observations cast doubts on this, and other scenarios have been proposed (see, e.g., \citealt{vanderklis2001,homan2002,homan2007,church2012}). In addition to motion along the tracks these sources show another type of motion, usually on longer timescales (days to weeks), where the tracks can shift, primarily in the HID. This has been referred to as secular motion/shifts/changes/variations. In most cases these shifts are small and do not lead to significant changes in the shapes of the tracks. However, a few sources have been known to exhibit much stronger secular changes where the tracks shift and change shape radically in both the CD and HID.

A breakthrough in our understanding of the Z/atoll phenomenology and the nature of secular evolution came with the transient NS-LMXB \src, which over the course of its 2006--2007 outburst evolved through all subclasses of low-magnetic-field NS-LMXBs---from a Cyg-like Z source at the highest luminosities to a Sco-like one, followed by a phase in the atoll source soft state (during which type~I X-ray bursts were seen), and ending with a transition to the atoll source hard state before returning to quiescence \citep{lin2009a,homan2010}. \citet{lin2009a} performed a detailed spectral analysis of the entire outburst, using data from the \textit{Rossi X-ray Timing Explorer} (\textit{RXTE}), and their results indicate that changes in the mass accretion rate were responsible for the evolution of the source. \citet{homan2010}---hereafter referred to as \citetalias{homan2010}---further argue that the observed behavior of the source implies that differences in the mass accretion rate can alone explain the existence of the various NS-LMXB subclasses, and that it is not necessary to invoke differences in other parameters, such as the magnetic field of the neutron star \citep{hasinger1989,psaltis1995} or the inclination angle of the system \citep{kuulkers1995}.

The main goal of this paper is to study to what extent (if any) evolution similar to that of \src\ is observed in other sources. Comparisons of \src's evolution at the low-luminosity end with atoll sources have been made in \citet{lin2009a}, \citetalias{homan2010}, and \citet{munoz2014}, showing that there the source behaved similar to other low-luminosity (atoll) NS-LMXBs; in this paper we focus mainly on the high-luminosity (Z source) portion of \src's evolution. To this end we performed a detailed analysis of the CDs and HIDs of three luminous neutron star X-ray binaries---\cyg, \cir, and \gx---using all available \xte\ Proportional Counter Array (PCA) data. Similar to \src, these sources are known to have shown strong secular evolution \citep[e.g.,][]{oosterbroek1995,kuulkers1996a,wijnands1997,shirey1999a,schnerr2003}. Since we are mainly interested in sources that cross subclass boundaries, we do not include in our study persistent Z sources that show only mild secular shifts and no significant changes in the shapes of their tracks. Although small subsets of the \xte\ data sets for \cyg, \cir, and \gx\ have been studied in several papers---usually with an emphasis on spectroscopy and/or timing analysis---we are not aware of any papers presenting a comprehensive study of the secular evolution of any of these sources using a large amount of \xte\ data, with the exception of \citet{schnerr2003}, who studied a large set of observations of \gx\ made in 1998 (see discussion in Section~\ref{sec:gx_13+1}). However, since then the amount of \gx\ data in the \xte\ archive has grown considerably. We also note that Shirey and collaborators studied CDs and HIDs of \cir\ using a number of observations made in 1996 and 1997 in a series of papers (\citealt{shirey1996,shirey1998,shirey1999a}; see discussion in Section~\ref{sec:cir_x-1}), but these observations represent only a small fraction of the currently available \xte\ data for this source. In addition to our analysis of \cyg, \cir, and \gx\ we also perform some reanalysis of \src\ in order to ensure complete consistency with the analysis of the other three sources and to facilitate comparisons between them and \src. We note that although \cir\ may be a rare low-magnetic-field neutron star high-mass X-ray binary (as further discussed in Section~\ref{sec:cir_x-1}) we will for simplicity in general refer to all four sources studied in this paper as NS-LMXBs.

The structure of the paper is as follows. In Section~\ref{sec:analysis} we describe the general data analysis steps common to all four sources, but leave the description of analysis specific to each source to Section~\ref{sec:results}. In Section~\ref{sec:1701} we briefly describe our analysis and present our results for \src. This analysis largely follows that previously performed by \citetalias{homan2010} (and includes creating a sequence of CD/HID tracks that shows the secular evolution of the source); we therefore mostly refer to that paper for details. In this section we also tie this source's behavior to the various source states/branches generally seen in the CDs and HIDs of NS-LMXBs. In Sections \ref{sec:cyg_x-2}--\ref{sec:gx_13+1} we describe our analysis and present our results for \cyg, \cir, and \gx, respectively. For each source we construct a sequence of CD/HID tracks (analogous to the one for \src) that illustrates its secular evolution; we also give a brief background on each of these three sources. In Section~\ref{sec:discussion} we discuss our results, and in Section~\ref{sec:summary} we give a summary of our results and conclusions.

\section{Data Analysis}\label{sec:analysis}

\subsection{Data Extraction}\label{sec:extraction}

We used the PCA data from all pointed \xte\ observations of \src, \cyg, \cir, and \gx\ obtained during the lifetime of the mission. In Table~\ref{tab:source_sample} we list the total amount of useful data (i.e., data remaining after the removal of X-ray bursts and bad data of any kind) for each of these sources. In the case of \cir, however, the quoted exposure time includes data affected by local absorption; see Section~\ref{sec:cir_x-1}. Because of the large number of observations (over 2300 individual ObsIDs), most of the data analysis steps described below were automated.

\begin{deluxetable}{lc}[t]
\tablewidth{8.6cm}
\tablecaption{Total Exposure Times for Analyzed Sources \label{tab:source_sample}}
\tablehead{\colhead{\hspace{-1.24cm}Source} & \colhead{Exp.\ Time (Ms)}}
\startdata
\src\		&	2.71 \bigstrut[t] \\[0.6ex]
\cyg\		&	2.28 \\[0.6ex]
\cir\		&	2.57 \\[0.6ex]	
\gx\		&	0.58 \bigstrut[b]
\enddata
\end{deluxetable}

The data were analyzed using HEASOFT, versions 6.9--6.12, as well as locally developed software. We used the \textit{Standard-2} data, extracting counts in \mbox{16 s} time bins, combined from all three xenon layers, from all active Proportional Counter Units (PCUs) at any given time. The data were filtered following the recommendations of the \xte\ Guest Observer Facility (GOF);\footnote[4]{See \url{http://heasarc.gsfc.nasa.gov/docs/xte/xhp_proc_analysis.html}.} this included the removal of data around PCU voltage breakdown events (but only from the relevant PCU in each case). We also corrected for dead time using the standard procedure recommended by the \xte\ GOF and subtracted background using the faint or bright background model as appropriate in each observation, based on the average count rate during the observation after exclusion of any type I X-ray bursts.

\subsection{Burst and Particle Flare Removal}\label{sec:removal}

We removed all data obtained during type I X-ray bursts; to identify bursts in an automated fashion we used a method similar to the one used by \citet{remillard2006b}. We note that although some bursts observed from \cyg\ and \cir\ do not show cooling along their tails---as is usually observed in type~I X-ray bursts---the origin of these faint bursts is nevertheless very likely thermonuclear \citep{linares2011} and they were removed.

PCUs~0 and 1 lost their propane layers in 2000 and 2006, respectively, diminishing their ability to reject events due to particle background. Based on a comparison with data from PCU~2---which did not suffer from such particle flares and was nearly always on---these events were identified and subsequently removed from the data. We note that since PCU~2 data exist for all times during which PCU~0/1 data were removed, we did not exclude data from any given point in time entirely, but simply reduced the number of PCUs contributing to a particular time bin in these cases.

\subsection{Response Correction}\label{sec:correction}

From the \textit{Standard-2} data we extracted count rates in several different energy bands and used those to calculate colors for CDs and HIDs. The response of the \xte\ PCUs evolved over the duration of the mission due to several factors, and these changes in response must be corrected for before data across the lifetime of the mission can be combined into a single CD or HID for a given source. For this correction we used observations of the Crab Nebula, which was observed on average $\sim$3 times per month throughout the mission. For each Crab observation we extracted the average count rate in each of the energy bands of interest for each active PCU, using the \textit{Standard-2} data. We then fitted the mission-long light curve (where each data point represented the average count rate during a single observation) for each PCU in each energy band with a piecewise-linear function. We (somewhat arbitrarily) chose MJD 50800 (1997 December 18) as our reference epoch and normalized all data points to that date, first for each PCU individually, and then in each case the count rates from PCUs 0, 1, 3, and 4 to that of PCU~2.

Recently it has become apparent that X-ray emission from the Crab Nebula shows significant intrinsic variability. \citet{wilson-hodge2011} show that from 2001 to 2010 the \xte\ PCA count rate from the Crab (after correcting for changes in response) varied rather irregularly by several percent. The (response-corrected) PCU~2 variability is $\sim$5\% in the \mbox{2--15 keV} band and $\sim$8\% in the \mbox{15--50 keV} band. This can compromise our correction for the variation in the PCA response. Comparing our mission-long Crab light curves to those of \citet{wilson-hodge2011} we can estimate the magnitude of the effect on our derived rates and colors. The relative amplitude of the Crab count rate fluctuations gradually increases with energy, but otherwise the variability behaves in more or less the same way in the different energy bands. Due to this energy dependence of the variability it should have some effect on the colors. However, given the relatively small difference in the strength of the variability between adjacent energy bands, we expect that this variability will largely cancel out in the colors. We estimate that shifts in our colors due to the Crab variability are at most $\sim$1\%--2\% and in most cases significantly less; we therefore expect the influence of this on our results to be negligible (cf.\ discussion about uncertainty due to counting statistics in Section~\ref{sec:rebinning}). The intensity is affected more strongly; we expect that shifts there can possibly be as high as $\sim$6\% in the worst case. However, they are generally much smaller---probably less than $\sim$4\% in almost all cases and typically in the 0\%--3\% range---and we expect the effect of this on our results to in general be negligible as well.

\subsection{Construction of CDs and HIDs}\label{sec:rebinning}

For the creation of CDs (i.e., hard color versus soft color) and HIDs (hard color versus intensity) we used color definitions similar to those used in \citetalias{homan2010}: we defined our soft color as the net counts in the \mbox{4.0--7.3 keV} band divided by those in the \mbox{2.4--4.0 keV} band, and our hard color as net counts in the \mbox{9.8--18.2 keV} band divided by those in the \mbox{7.3--9.8 keV} band. The intensity we used for the HID was the net count rate per PCU in the \mbox{2--60 keV} band. Before creating our CDs and HIDs we combined the counts from all active PCUs for each \mbox{16 s} time bin (after performing the corrections and filtering described above) and then rebinned the data in a given observation in order to maintain a more uniform size in the error bars across different values in count rate. We set a minimum of 16,000~counts in the \mbox{2--60 keV} band (after background subtraction) for each rebinned data point. The rebinning was done in an adaptive/dynamic fashion---i.e., we in general did not use a single value for the time binning factor over an entire observation, but instead allowed the factor to vary (while imposing our counts limit) to adapt to a possibly varying count rate. In a few cases we applied a larger counts limit; these will be mentioned explicitly. In some instances entire observations did not contain enough counts to reach the 16,000~counts minimum; in those cases we created a single data point from the whole observation if the total number of counts was larger than 10,000, but excluded it from further consideration if it had fewer counts. (Any exceptions to this will be explicitly mentioned.) We note that for most observations of the four bright sources studied here no rebinning was necessary, since each \mbox{16 s} time bin usually contained more than the minimum 16,000~counts. With this minimum counts limit the uncertainty due to counting statistics is at most $\sim$2\% in the soft color and at most $\sim$3\% in the hard color (often much less); the uncertainty in the intensity is always less than 1\%.

\begin{figure}[t]
\centerline{\includegraphics[width=8.6cm,trim=0 0 0 -15]{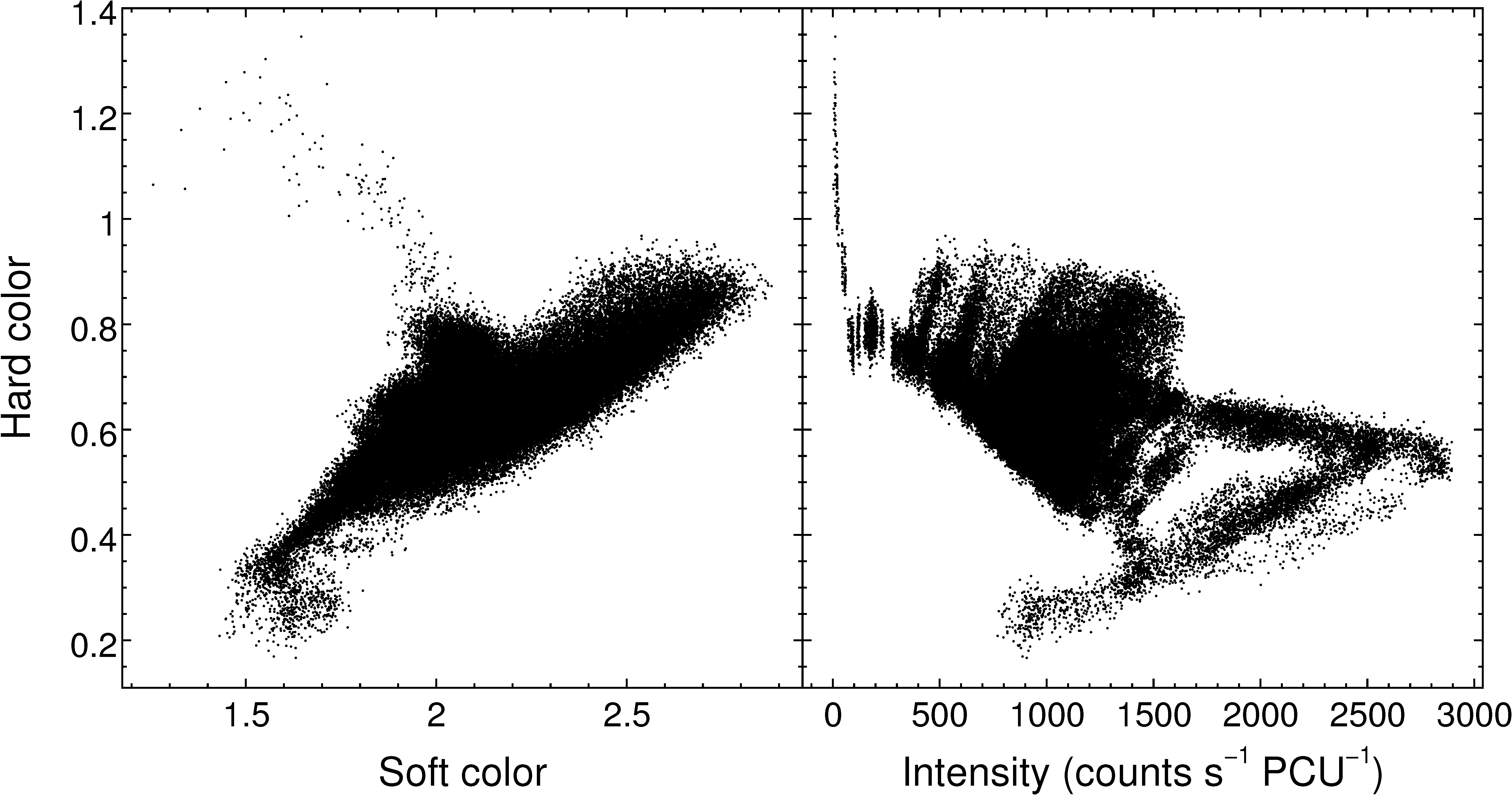}}
\caption{CD and HID representing the entire \xte\ PCA data set of \src.}\label{fig:1701_all_data}
\end{figure}

\section{Results}\label{sec:results}

\subsection{\src}\label{sec:1701}

\src\ was discovered with the \xte\ All-Sky Monitor (ASM) on 2006 January 18 as the source was entering outburst \citep{remillard2006a}. The first \xte\ PCA observation took place on 2006 January 19, and the source was subsequently observed with the PCA on average \mbox{$\sim$1.5--2} times per day for the remainder of the 19 month outburst, apart from a $\sim$50~day period in late 2006 and early 2007 during which the source could not be observed due to proximity to the Sun. The source returned to quiescence in early 2007 August and has since remained inactive apart from occasional low-level flaring (up to \mbox{$\sim$$10^{35}\textrm{ erg s}^{-1}$}), which has been observed by \sw\ and \xmm\ \citep{fridriksson2010,fridriksson2011}. As in \citetalias{homan2010}, dates during the outburst will in this paper be referred to as days since the start of 2006 January 19 (MJD 53754.0).

\begin{figure*}[t]
\centering
\includegraphics[height=11.9cm]{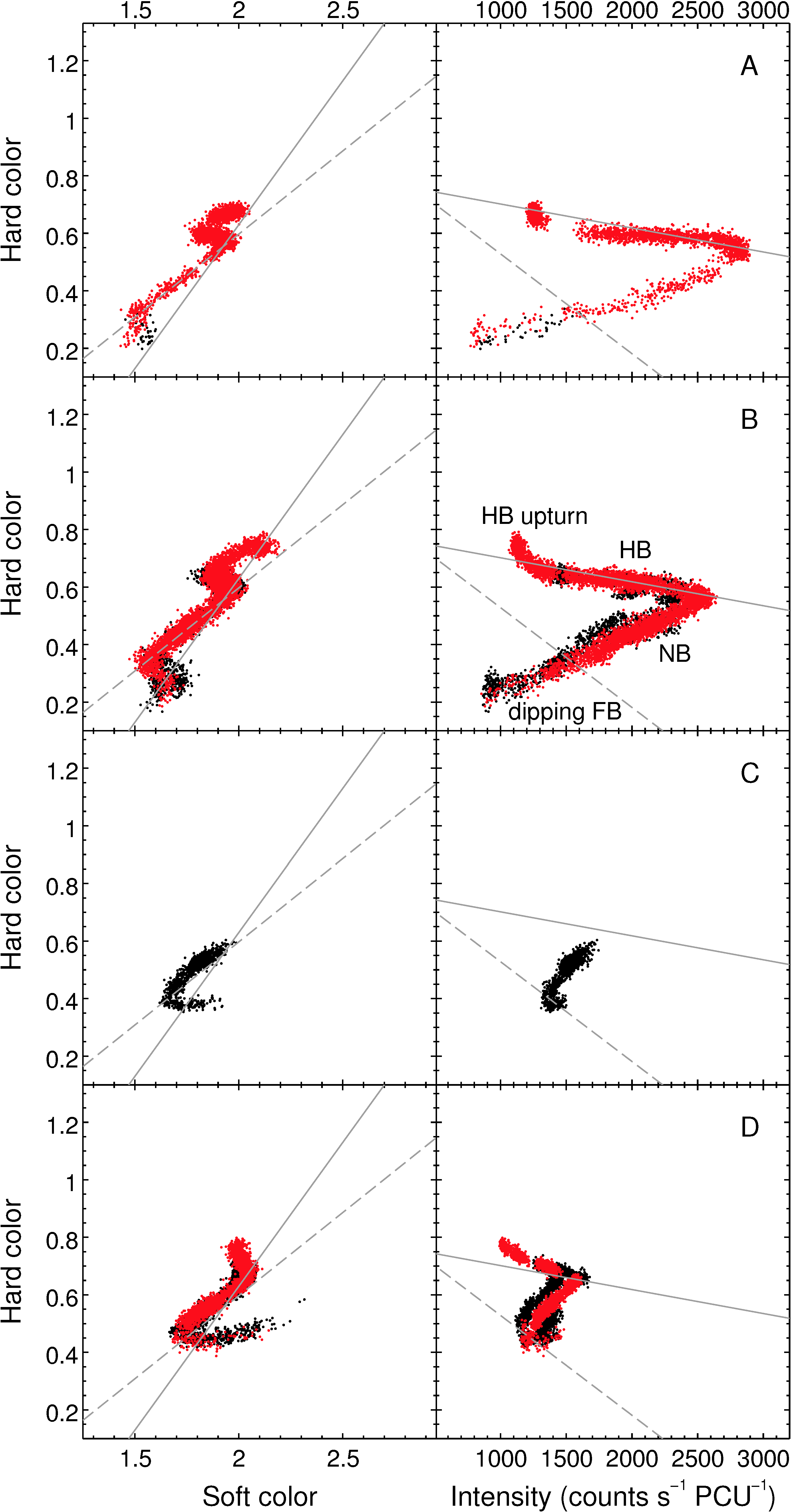}
\hspace{-0.04cm}
\includegraphics[height=11.9cm]{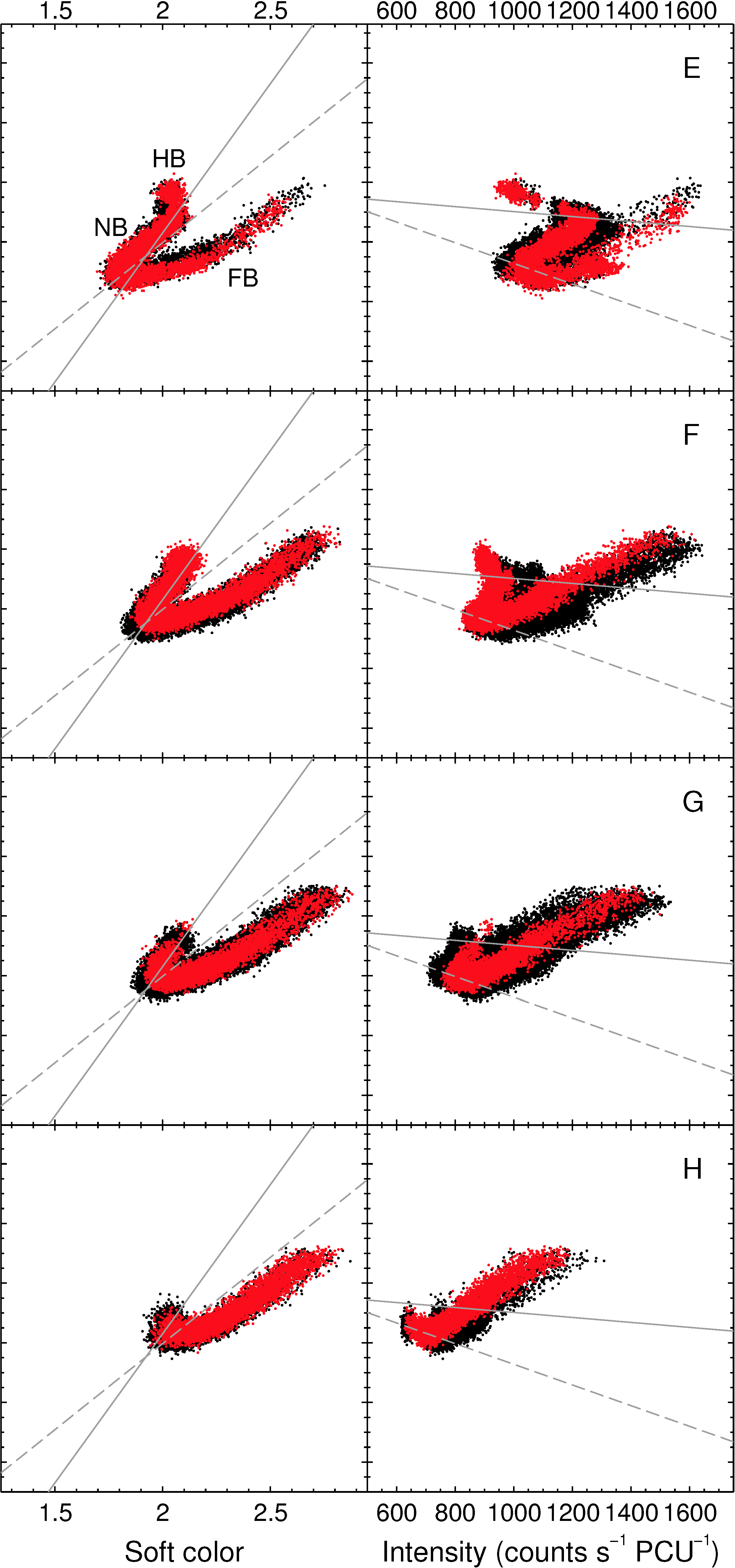}
\hspace{-0.04cm}
\includegraphics[height=11.9cm]{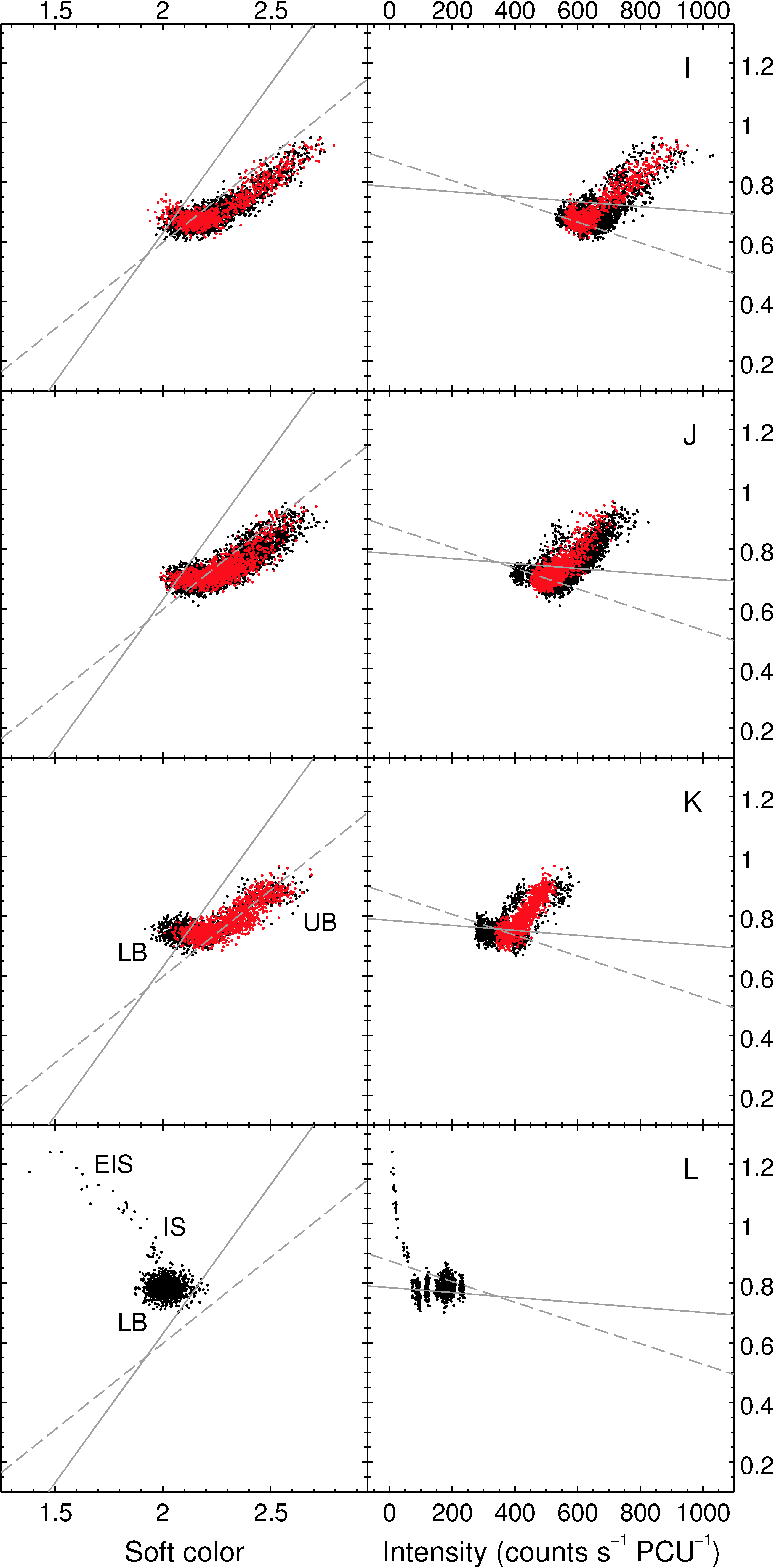}
\caption{CDs and HIDs for \src, showing the secular evolution of the source. Each of the 12 panels (consisting of a CD on the left and an HID on the right) corresponds to a particular selection of data from the entire \xte\ PCA data set for the source; red data points correspond to a particular subset (exhibiting minimal secular shifts) within a given selection (see Table~\ref{tab:selections}). Particular source states/branches are indicated in a few of the panels (see text for definitions). The dashed and solid lines show the approximate paths followed by the NB/FB and HB/NB vertices, respectively, as the tracks evolve. Data are binned to a minimum of 16,000~counts per data point, except for the IS and EIS data in selection L, which are binned to a minimum of 64,000~counts. Note the change in the intensity scale between the different HID columns. We also note that essentially the same tracks (with only small modifications here) were presented in slightly different form in Figures 3 and 4 in \citetalias{homan2010}.}\label{fig:1701_sequence}
\end{figure*}

The combined CD and HID for the entire 2006--2007 outburst of \src\ are shown in Figure~\ref{fig:1701_all_data}; the diagrams clearly illustrate that the source tracks exhibited strong secular motion during the outburst. As in \citetalias{homan2010} we divided the observations of the source into 12 subsets---hereafter referred to as selections---which we label A--L. In Figure~\ref{fig:1701_sequence} we show the CDs and HIDs for all 12 selections, which illustrate the secular evolution of the source during the outburst. We label the source states/branches of a few representative tracks: a Cyg-like Z track (selection B), a close-to Sco-like Z track (E), and two atoll-like tracks (K and L). For the Z-like tracks these are the horizontal branch (HB), the HB upturn, the normal branch (NB), the flaring branch (FB), and the dipping FB. For the atoll-like tracks these are the lower and upper banana branches (LB and UB), the island state (IS), and the extreme island state (EIS). The banana branch, IS, and EIS are also referred to as the soft, intermediate/transitional, and hard state, respectively. (We refer to \citetalias{homan2010} for examples of CDs/HIDs of several Cyg/Sco-like Z sources and atoll sources and a comparison of those with the CD/HID tracks of \src.)

The 12 data selections mostly correspond to particular ranges in the low-energy (\mbox{2.0--2.9 keV}) count rate, which \citetalias{homan2010} found to closely trace the secular changes during most of the outburst. This tracing of the secular evolution with the low-energy count rate breaks down when the source enters the HB or a dipping FB. For days~0--28 time-based selections (A and B) were therefore used, rather than ones based on the low-energy count rate. In addition, data from several HB excursions taking place after day~28 had to be moved to the same selection as neighboring NB data in the low-energy light curve. The time and count rate intervals we used are given in Table~\ref{tab:selections}. Figure~2 in \citetalias{homan2010} shows a low-energy light curve of the entire outburst, indicating the count rate intervals used there (which are for the most part equivalent to ours) and the HB data moved between selections.

\begin{deluxetable}{lcc}[]
\tablewidth{8.6cm}
\tablecaption{Time or Count Rate Intervals Used for Data Selections \label{tab:selections}}
\tablehead{\colhead{\hspace{-0.17cm}Selection} & \colhead{Full Interval} & \colhead{Subinterval} \\[0.4ex]
 & \colhead{(counts s$^{-1}$ PCU$^{-1}$)\tablenotemark{a}} & \colhead{(counts s$^{-1}$ PCU$^{-1}$)\tablenotemark{a}}}
\startdata
A & Days 13.5--20.5 & Days 13.5--19.5 \bigstrut[t] \\[0.6ex]
B & Days 0--13.5 and 20.5--28 & Days 2--13.5 \\[0.6ex]
C & 106--114/Days 28--32.5 & \nodata \\[0.6ex]
D & 83--106 & 89--97 \\[0.6ex]
E & 66.5--83 & 72.5--78 \\[0.6ex]
F & 56--66.5 & 56--58.36 \\[0.6ex]
G & 46--56 & 50--52 \\[0.6ex]
H & 38.5--46 & 39.6--43 \\[0.6ex]
I & 31--38.5 & 33.5--35.5 \\[0.6ex]
J & 23.5--31 & 26--28.5 \\[0.6ex]
K & 15--23.5 & 20--22.5 \\[0.6ex]
L & 0.4--15/Days 550--564 & \nodata \bigstrut[b]
\enddata
\tablenotetext{a}{Count rates are in the \mbox{2.0--2.9 keV} band. Time intervals refer to days since MJD 53754.0.}
\tablecomments{}
\end{deluxetable}

Within most of the selections some secular motion is still evident, mostly in the HID. We therefore in 10 of the 12 selections color red the data points from a subset of the observations used in each case, to show tracks with minimal secular shifts; the subintervals in time or count rate defining these subsets are given in Table~\ref{tab:selections}. We also show, in both the CDs and HIDs, two straight lines that the NB/FB (lower) and HB/NB (upper) vertices of the tracks are seen to follow closely (as pointed out by \citealt{lin2009a}). We note that selections A--L do not represent a strict monotonic progression in time (although overall the evolution in time was from A to L); the source moved back and forth between selections during the outburst, mostly within the range F--H (see Figure~2 in \citetalias{homan2010}). It is clear from the sequence in Figure~\ref{fig:1701_sequence} that going from panel A to L the tracks smoothly evolve in shape from Cyg-like Z to Sco-like Z to atoll tracks as the upper and lower vertices both move monotonically along the vertex lines to higher color values and the overall intensity of the tracks decreases.

\begin{figure}[t]
\centerline{\includegraphics[width=8.6cm, trim = 0 0 0 -15]{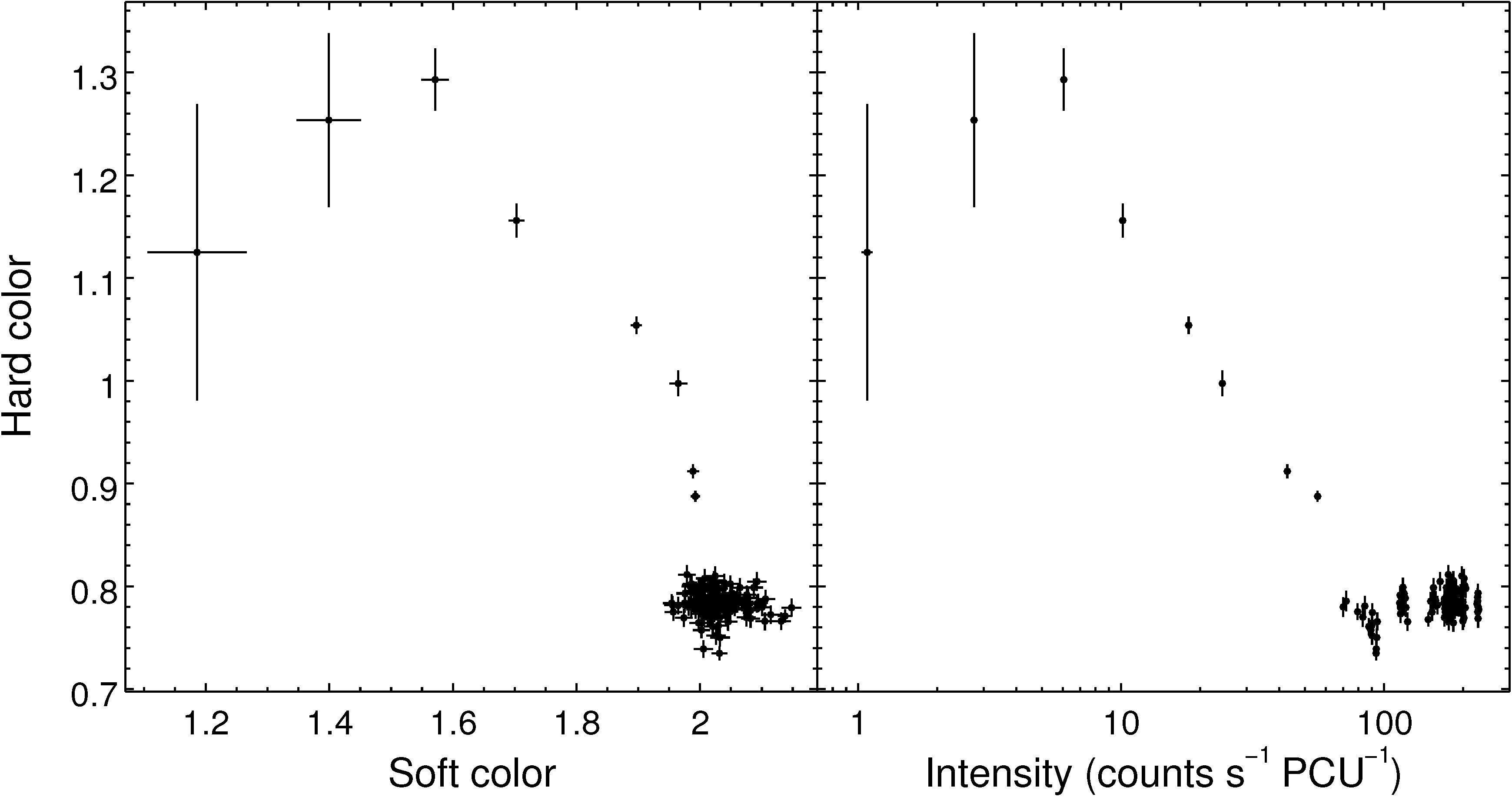}}
\caption{Alternative version of panel L in Figure~\ref{fig:1701_sequence} with the data binned into fewer groups and Galactic background emission subtracted. Data in the LB (i.e, intensity \mbox{$\gtrsim$65 counts s$^{-1}$ PCU$^{-1}$} and hard color $\lesssim$0.85) are binned to a minimum of 128,000~counts per data point; data in the IS and EIS are binned to approximately one data point per day. The data represented by the leftmost point were not included in Figure~\ref{fig:1701_sequence}.}\label{fig:1701_atoll}
\end{figure}

After day~566 (as the outburst was ending) the count rate from the source reached a roughly constant level of \mbox{$\sim$2 counts s$^{-1}$ PCU$^{-1}$} in the \mbox{2--60 keV} band (\mbox{$\sim$0.2 counts s$^{-1}$ PCU$^{-1}$} in the low-energy band). This residual flux can be attributed to diffuse Galactic background emission (see \citealt{fridriksson2010}). Figure~\ref{fig:1701_atoll} shows an alternative version of panel L in Figure~\ref{fig:1701_sequence} where we have binned the data more heavily and have subtracted this diffuse background emission based on data from observations made during the three weeks following the end of the outburst. In addition, we include in this version data from two observations made during days~564--566 (constituting the leftmost data point), which had a low-energy count rate of \mbox{$\sim$0.36 counts s$^{-1}$ PCU$^{-1}$}. We furthermore use a logarithmic scale for the horizontal axis in the HID so as to better illustrate the behavior of the source at the lowest count rates.

\subsection{\cyg}\label{sec:cyg_x-2}

Cygnus X-2 (\cyg) is one of the longest-known and most extensively studied X-ray binaries. \cyg\ was classified as a Z source by \citet{hasinger1989} based on \mbox{\textit{EXOSAT}} data and is the prototype of the Cyg-like subgroup of the persistent Z sources. However, as was clear already from pre-\xte\ data \cyg\ is unique among the six ``classic'' (persistent Galactic) Z sources in that it shows by far the strongest secular evolution \citep[e.g.,][]{kuulkers1996a,wijnands1997}.

\begin{figure}[b]
\centerline{\includegraphics[width=8.5cm]{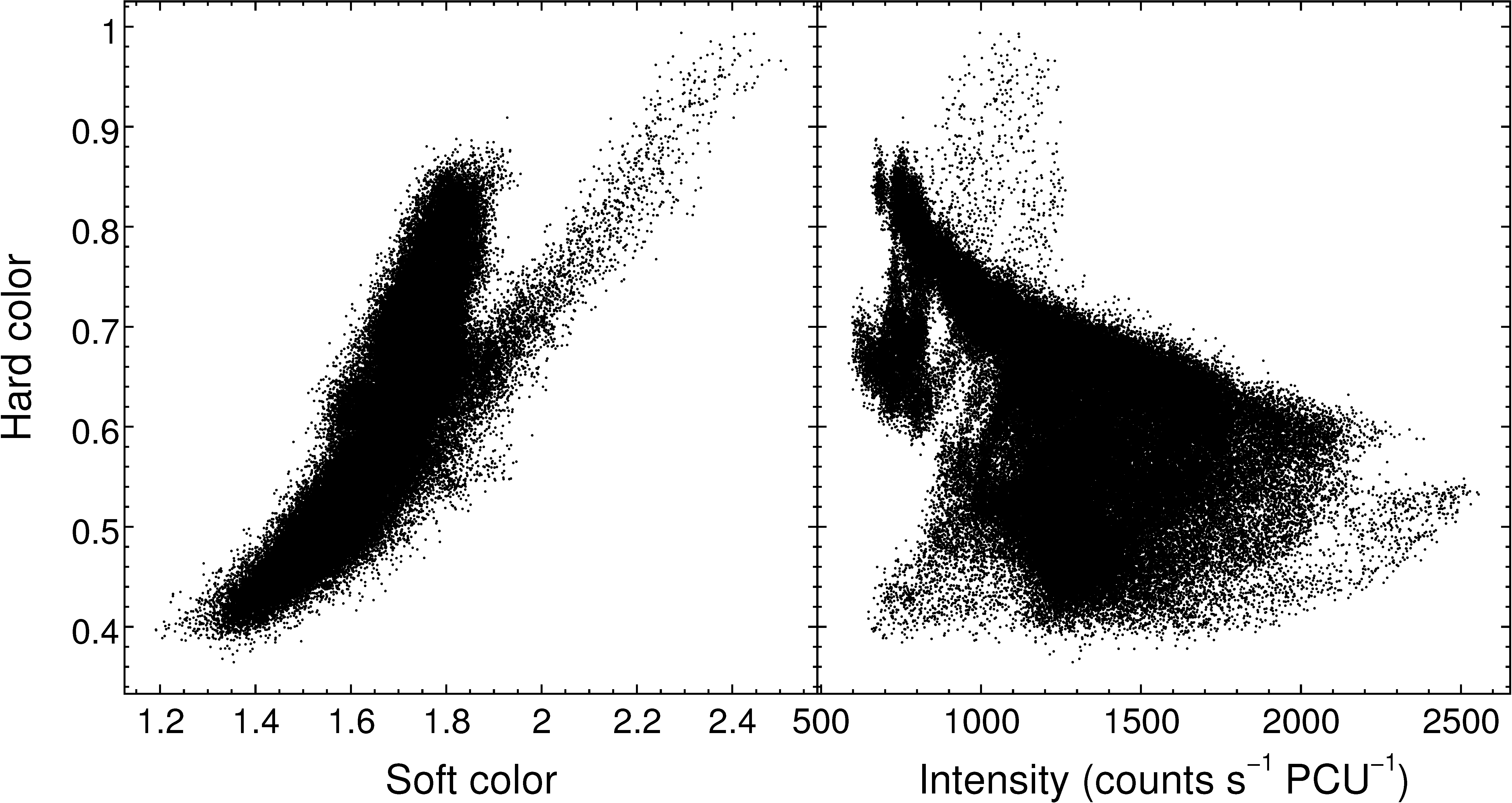}}
\caption{CD/HID representing the entire \xte\ PCA data set of \cyg.}\label{fig:cyg_x-2_all_data}
\end{figure}

\subsubsection{Analysis}\label{sec:cyg_x-2_analysis}

In Figure~\ref{fig:cyg_x-2_all_data} we show a CD/HID based on the entire \xte\ PCA data set for the source; strong variations in the shape and location of tracks are readily apparent. This combined CD/HID for \cyg\ has strong similarities to the one for \src\ (Figure~\ref{fig:1701_all_data}). However, the fragmented nature of the data set (obtained over a period of 15~years) forces us to analyze the data in a manner different from the \src\ analysis.

\begin{figure*}[t]
\centering
\includegraphics[height=14.75cm]{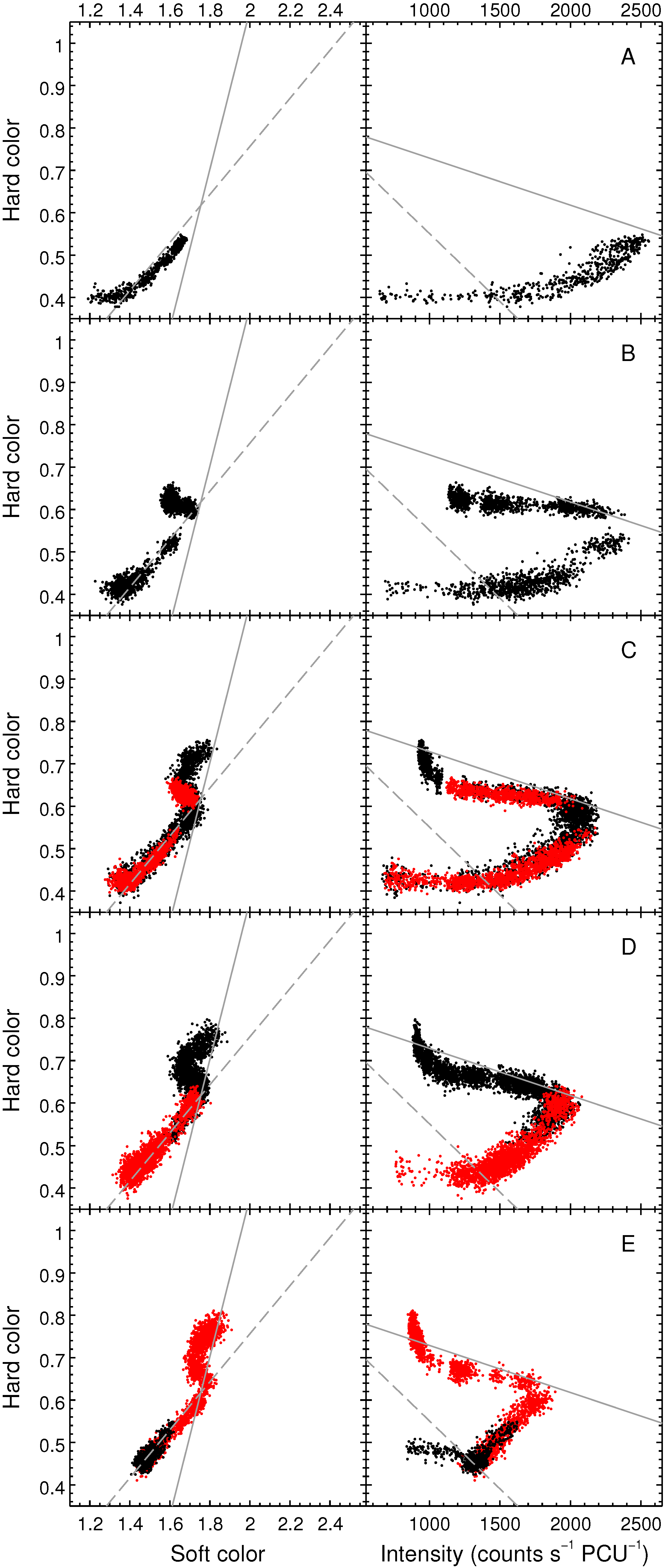}
\hspace{-0.055cm}
\includegraphics[height=14.75cm]{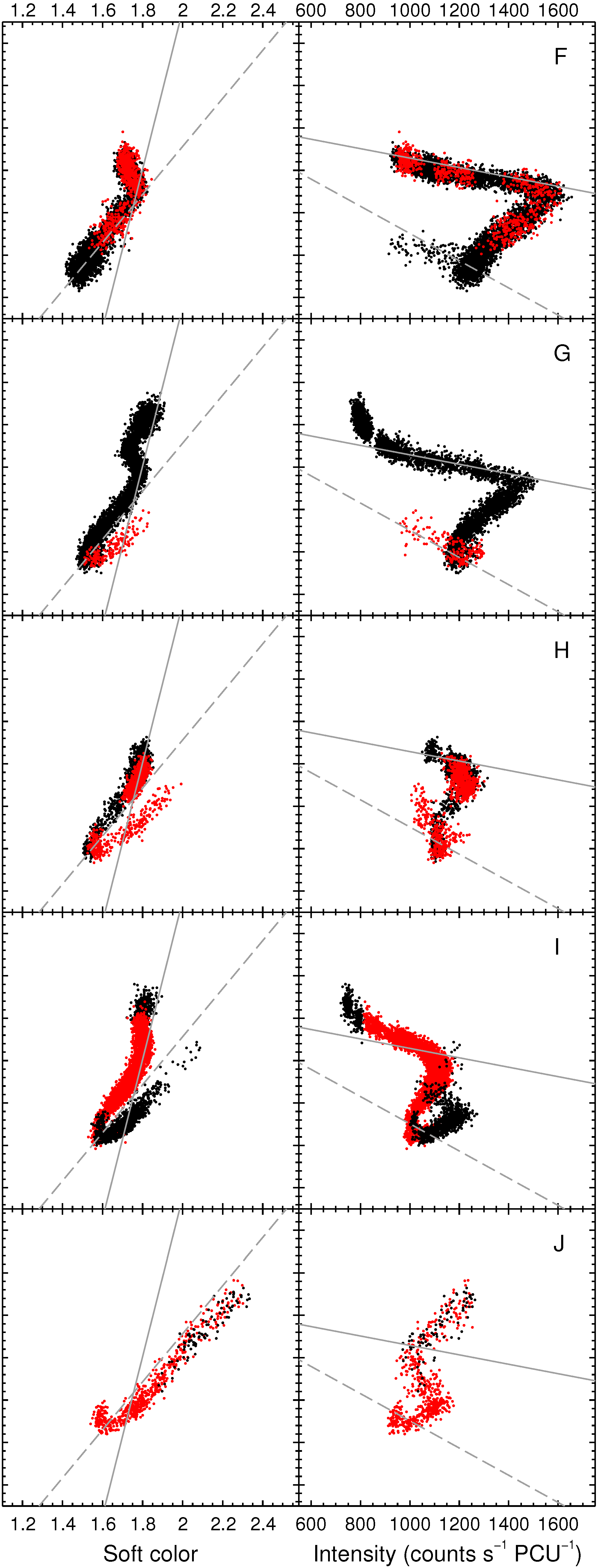}
\hspace{-0.055cm}
\includegraphics[height=14.75cm]{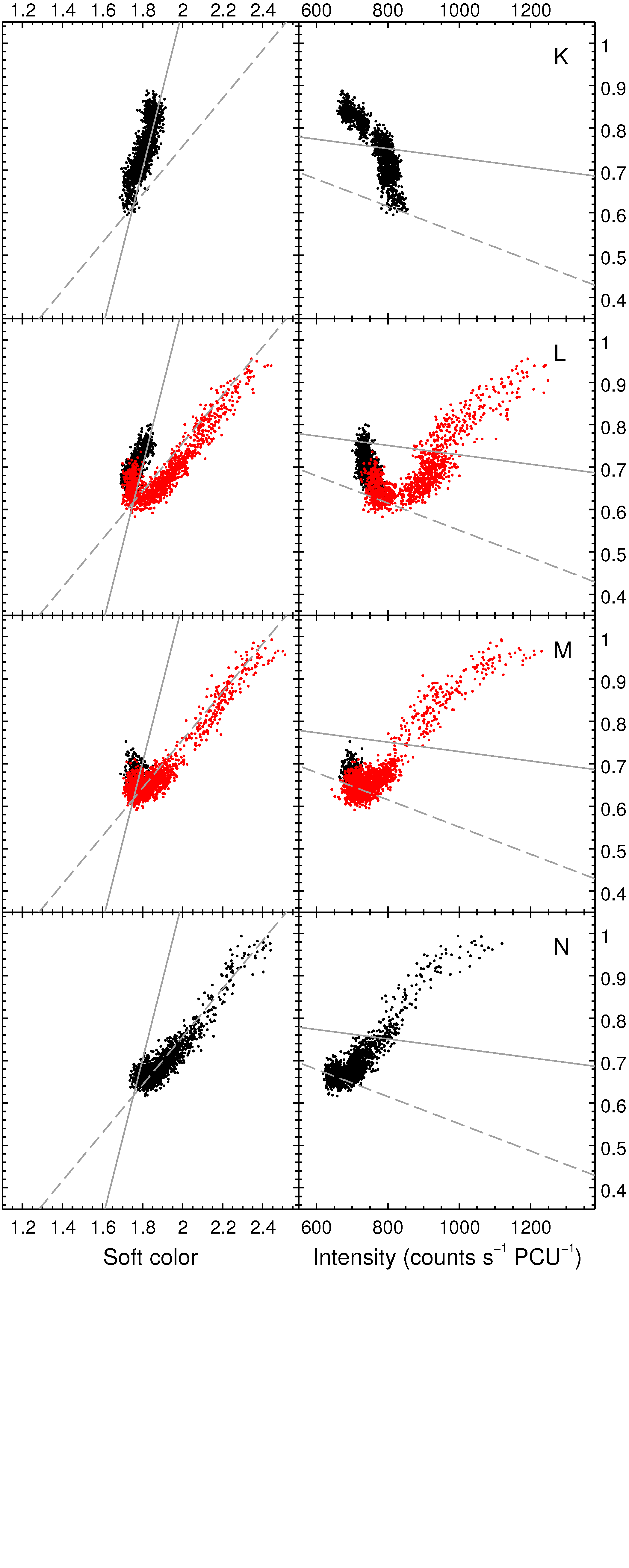}
\caption{A sequence of CDs and HIDs for \cyg\ illustrating the secular evolution of the source. Table~\ref{tab:cyg_x-2_obs} lists the data used in each panel. In general, a panel combines data from observations widely separated in time (by months or years); exceptions to this are panels A, B, K, and N. In each of the other panels, a particular subset of the data---obtained over a period of at most a few days---is shown in red. The dashed and solid lines show the approximate paths followed by the lower and upper vertices, respectively. Note the change in the intensity scale between the different HID columns.}\label{fig:cyg_x-2_sequence}
\end{figure*}

For this analysis we considered all 591 individual ObsIDs for pointed observations of \cyg\ made during the \xte\ mission (13 of which contained no useful data). Often multiple observations were fairly densely clustered together in time. Going chronologically through the data set the source jumps erratically back and forth around the CD/HID, in any given observation usually tracing out only short partial track segments. As a first step we went through the entire data set in time order, at any point combining into a single CD/HID track as many consecutive ObsIDs as possible without introducing clearly noticable secular shifts. We refer to such clusters of data (containing all data from a certain time interval) as \textit{subsets}. In a few cases significant secular motion took place during a single ObsID, requiring the observation to be split up between two different subsets. To identify secular motion we examined the tracks not only in the CD and HID, but also in a diagram of soft color versus intensity (which we refer to as an SID). The SID frequently yielded useful extra information/constraints. The result of this process was $\sim$300 subsets, which span periods as long as several days, although the vast majority is less than a day in length.

Considering only individual subsets results in a few tracks that seem to a large extent complete, but most of the subsets consist of shorter segments (i.e., incomplete tracks). By an ``incomplete'' track we mean that had the source stayed long enough at a particular stage in its secular progression (and had our observational coverage of the source been sufficiently comprehensive during that time) we expect that the source would have traced out a fuller (i.e., continuous and possibly more extended) track. It is important to note that this ``completeness'' is a function of location in the secular progression, as not all branches of the Z track are present at every stage of the secular evolution, and that there is always some uncertainty regarding what constitutes a ``complete'' track.

Fortunately, it was in general possible to combine subsets from various times throughout the \xte\ mission to form more complete tracks than are available from individual subsets. Usually, the most complete individual-subset tracks served as a foundation on which these combined tracks were built. To guide us in this process we took advantage of overlapping track segments and made sure that they lined up in all three diagrams---the CD, HID, and SID---which together provided a fairly stringent criterion for the appropriateness of combining particular segments.

\setlength\bigstrutjot{3pt}
\tabletypesize{\scriptsize}

\begin{deluxetable*}{clclc}[]
\tablewidth{18.0cm}
\tablecolumns{5}
\tablecaption{Time Intervals and Observations Used to Create \cyg\ Tracks \label{tab:cyg_x-2_obs}}
\tablehead{ & & \colhead{Interval Length} & & \colhead{Exp.\ Time} \\[0.3ex]
\colhead{\hspace{-0.2cm}Panel} & \colhead{\hspace{-0.55cm}Time Interval (MJD)} & \colhead{days (hr)\tablenotemark{a}} & \colhead{\hspace{-4.03cm}ObsIDs} & \colhead{ks (hr)\tablenotemark{b}}}
\startdata
A			       & 50316.592--50316.764 	     & 0.17 (4.1)\phn\phn			& 10063-[09:10]-01-00 				   		   	& 9.0 (2.5) \bigstrut[b] \\
\hline
B			       & 51697.550--51698.705 	     & 1.15 (27.7)\phn				& 40019-04-05-[00:10]  						   	& 33.9 (9.4)\phn \bigstrut \\
\hline
C			       & 50996.450--51000.593 (R)   & 4.14 (99.4)\phn				& 30418-01-[(01:05)-00,02-01]       	          		   	& 37.4 (10.4) \bigstrut[t] \\[0.4ex]
			       & 51561.344--51561.517 	     & 0.17 (4.1)\phn\phn			& 40017-02-19-00 							   	& 10.1 (2.8)\phn \\[0.4ex]
			       & 53138.767--53138.855 	     & 0.09 (2.1)\phn\phn			& 90030-01-16-00 							   	& 5.2 (1.4) \\[0.4ex]
			       & 53788.496--53788.577 	     & 0.08 (2.0)\phn\phn			& 91009-01-42-[00:01] 						   	& 2.7 (0.7) \\ [0.4ex]
			       & 54649.890--54650.626 	     & 0.74 (17.6)\phn				& 93443-01-01-[02,14:21] 						   	& 23.6 (6.6)\phn \bigstrut[b] \\
\hline			     
D			       & 51009.488--51009.725	     & 0.24 (5.7)\phn\phn			& 30046-01-01-00							   	& 12.9 (3.6)\phn \bigstrut[t] \\[0.4ex]
			       & 51349.631--51349.669	     & 0.04 (0.9)\phn\phn			& 40017-02-09-01							   	& 3.3 (0.9) \\[0.4ex]
			       & 53079.687--53079.785	     & 0.10 (2.3)\phn\phn 			& 90030-01-04-00							   	& 5.2 (1.4) \\[0.4ex]
			       & 53222.440--53222.532	     & 0.09 (2.2)\phn\phn			& 90030-01-33-[00:01]						   	& 4.8 (1.3) \\[0.4ex]
			       & 53286.420--53286.512	     & 0.09 (2.2)\phn\phn	 		& 90030-01-46-00							   	& 5.3 (1.5) \\[0.4ex]
			       & 53291.340--53291.433	     & 0.09 (2.2)\phn\phn			& 90030-01-47-00							   	& 5.4 (1.5) \\[0.4ex]
			       & 54007.232--54007.331	     & 0.10 (2.4)\phn\phn	 		& 92039-01-15-00							   	& 5.7 (1.6) \\[0.4ex]
			       & 54009.195--54010.341	     & 1.15 (27.5)\phn				& 92039-01-[17,18]-00						   	& 8.2 (2.3) \\[0.4ex]
			       & 54452.324--54452.554	     & 0.23 (5.5)\phn\phn			& 90022-08-04-[00:01]						   	& 9.5 (2.6) \\[0.4ex]
			       & 54648.066--54649.329 (R)   & 1.26 (30.3)\phn	 			& 93443-01-01-[00,000,03:13]					   	& 37.2 (10.3) \bigstrut[b] \\
\hline			     
E			       & 50168.900--50169.731 (R)   & 0.83 (20.0)\phn				& 10066-01-01-[00,000,001]				   	   	& 37.3 (10.4) \bigstrut[t] \\[0.4ex]
			       & 51048.360--51048.599	     & 0.24 (5.7)\phn\phn			& 30046-01-07-00						   	   	& 13.4 (3.7)\phn \bigstrut[b] \\
\hline
F			       & 51055.495--51055.782	     & 0.29 (6.9)\phn\phn			& 30046-01-08-00							   	& 13.4 (3.7)\phn \bigstrut[t] \\[0.4ex]
			       & 51266.719--51267.126	     & 0.41 (9.8)\phn\phn	 		& 40017-02-05-[00:01]					   	   	& 15.9 (4.4)\phn \\[0.4ex]
			       & 51413.523--51413.746	     & 0.22 (5.4)\phn\phn			& 40017-02-12-00						   	   	& 13.0 (3.6)\phn \\[0.4ex]
			       & 52429.372--52429.908	     & 0.54 (12.9)\phn				& 70016-01-01-[02,04:06]					   	   	& 12.4 (3.4)\phn \\[0.4ex]
			       & 53511.868--53512.546	     & 0.68 (16.3)\phn				& 91010-01-01-[01,06:09]				   		   	& 29.9 (8.3)\phn \\[0.4ex]
			       & 53685.118--53685.193	     & 0.08 (1.8)\phn\phn	 		& 90030-01-82-00						   	   	& 4.3 (1.2) \\[0.4ex]
			       & 54395.861--54398.037 (R)   & 2.18 (52.2)\phn				& 92038-01-[08-(00:02),09-(00:02),10-(00:03)]	   	   	& 12.3 (3.4)\phn \bigstrut[b] \\
\hline
G			       & 51081.447--51081.725	     & 0.28 (6.7)\phn\phn			& 30046-01-12-00							   	& 14.3 (4.0)\phn \bigstrut[t] \\[0.4ex]
			       & 51444.401--51445.215	     & 0.81 (19.5)\phn				& 40019-04-[01-01G,01-02G,01-03,02-00,02-000]  	   	& 17.6 (4.9)\phn \\[0.4ex]
			       & 52533.929--52535.087	     & 1.16 (27.8)\phn				& 70015-02-[01-01G,02-00]				   	   	& 19.1 (5.3)\phn \\[0.4ex]
			       & 53311.026--53311.105	     & 0.08 (1.9)\phn\phn			& 90030-01-51-[00:01]					  	   	& 2.6 (0.7) \\[0.4ex]
			       & 53335.431--53335.724	     & 0.29 (7.0)\phn\phn			& 90022-08-01-00, 90030-01-56-[00:01]	   		   	& 8.2 (2.3) \\[0.4ex]
			       & 53375.032--53375.053	     & 0.02 (0.5)\phn\phn			& 90030-01-64-00						   	   	& 1.8 (0.5) \\[0.4ex]
			       & 53394.710--53394.812	     & 0.10 (2.5)\phn\phn			& 90030-01-68-00						   	   	& 6.3 (1.8) \\[0.4ex]
			       & 53488.223--53488.310	     & 0.09 (2.1)\phn\phn			& 91009-01-12-[00:01]					   	   	& 3.4 (0.9) \\[0.4ex]
			       & 53493.152--53493.227 (R)   & 0.07 (1.8)\phn\phn			& 91009-01-13-[00:01]					   	   	& 2.8 (0.8) \\[0.4ex]
			       & 53498.069--53498.142         & 0.07 (1.8)\phn\phn			& 91009-01-14-00						   	   	& 3.1 (0.9) \bigstrut[b] \\
\hline
H			       & 51539.192--51539.562 (R)   & 0.37 (8.9)\phn\phn			& 40017-02-18-[00:02]					   	   	& 14.4 (4.0)\phn \bigstrut[t] \\[0.4ex]
			       & 53119.069--53119.174         & 0.10 (2.5)\phn\phn	 		& 90030-01-12-00						   	   	& 6.8 (1.9) \\[0.4ex]
			       & 53330.721--53330.805         & 0.08 (2.0)\phn\phn	 		& 90030-01-55-[00:01]					   	   	& 2.9 (0.8) \\[0.4ex]
			       & 53347.497--53347.527         & 0.03 (0.7)\phn\phn 			& 90022-08-02-00						   	   	& 2.6 (0.7) \\[0.4ex]
			       & 53744.187--53744.278         & 0.09 (2.2)\phn\phn	 		& 90030-01-94-00						   	   	& 5.2 (1.4) \\[0.4ex]
			       & 54388.862--54389.040         & 0.18 (4.3)\phn\phn			& 92038-01-01-[00:02]					   	   	& 3.8 (1.1) \bigstrut[b] \\
\hline
I			       & 50629.832--50631.933 (R)   & 2.10 (50.4)\phn				& 20053-04-01-[00:04,06,010,020,030]\tablenotemark{c}  & 77.5 (21.5) \bigstrut[t] \\[0.4ex]
			       & 51029.751--51030.032	     & 0.28 (6.7)\phn\phn			& 30046-01-04-[00:01]					   		& 13.8 (3.8)\phn \\[0.4ex]
			       & 53266.795--53266.821 	     & 0.03 (0.6)\phn\phn			& 90030-01-42-00	 							& 2.3 (0.6) \\[0.4ex]
			       & 53758.990--53759.021	     & 0.03 (0.7)\phn\phn			& 90030-01-97-00						   		& 2.7 (0.7) \\[0.4ex]
			       & 54427.903--54428.970 	     & 1.07 (25.6)\phn				& 93082-02-04-[00:01] 						   	& 2.4 (0.7) \bigstrut[b] \\
\hline
J			       & 53123.996--53129.015 (R)   & 5.02 (120.5)				& 90030-01-[13:14]-00	   				   		& 8.8 (2.4) \bigstrut[t] \\[0.4ex]
			       & 53320.893--53320.934         & 0.04 (1.0)\phn\phn	 		& 90030-01-53-[00:01]				 	   		& 1.4 (0.4) \bigstrut[b] \\
\hline		     
K			       & 51528.887--51529.451         & 0.56 (13.5)\phn				& 40019-04-[03-01,04-(00:01),04-000]	 			& 22.8 (6.3)\phn \bigstrut \\   
\hline
L			       & 50719.393--50719.759 	      & 0.37 (8.8)\phn\phn			& 20057-01-01-[00,000]							& 20.2 (5.6)\phn \bigstrut[t] \\[0.4ex]
			       & 51536.473--51536.968 (R)    & 0.50 (11.9)\phn			& 40021-01-02-[00:04]\tablenotemark{d} 	   			& 19.2 (5.3)\phn \bigstrut[b] \\
\hline
M			       & 51535.886--51536.436 (R)   & 0.55 (13.2)\phn				& 40021-01-[01-01,02-00,02-000]\tablenotemark{d} 		& 26.7 (7.4)\phn \bigstrut[t] \\[0.4ex]
			       & 53202.754--53202.834 	      & 0.08 (1.9)\phn\phn			& 90030-01-29-00 						   		& 4.3 (1.2) \bigstrut[b] \\
\hline
N			       & 51535.010--51535.571          & 0.56 (13.5)\phn			& 40021-01-01-[00,000,02] 			 			& 25.8 (7.2)\phn \bigstrut
\enddata
\tablenotetext{a}{The interval length is shown in units of both days and hours.}
\tablenotetext{b}{The total exposure time is shown in units of both ks and hours.}
\tablenotetext{c}{As the source showed significant secular motion during observation 20053-04-01-04, only an early part of it was used.}
\tablenotetext{d}{The source showed clear secular motion during observation 40021-01-02-00. Data from the first two orbits were used in panel M and data from the third (and final) orbit was used in panel L.}
\tablecomments{Subsets colored in red in Figure~\ref{fig:cyg_x-2_sequence} are denoted by (R) in the Time Interval column. In the ObsIDs column a colon denotes a range.}
\end{deluxetable*}

In Figure~\ref{fig:cyg_x-2_sequence} we show a sequence of 14 tracks, chosen (from a larger set of tracks) to illustrate as clearly as possible the overall secular evolution exhibited by the source---both of the individual branches and the locations of the tracks---as they move through the CD/HID. As is the case for \src\ the lower and upper vertices approximately follow straight line paths in both the CD and HID as the tracks evolve and shift in the diagrams. We show illustrative lines in Figure~\ref{fig:cyg_x-2_sequence} and order the panels based on the vertex locations of the tracks, starting (panel A) at the highest intensities and lowest color values. 

The number of individual subsets in the tracks in the \cyg\ sequence ranges from 1 to 10. The subsets (i.e., time intervals) and ObsIDs used for each panel in Figure~\ref{fig:cyg_x-2_sequence} are listed in Table~\ref{tab:cyg_x-2_obs}. For the panels that consist of more than one data subset (i.e., all except A, B, K, and N), a single representative subset is plotted in red. The combined exposure time of the data shown in Figure~\ref{fig:cyg_x-2_sequence} is $\sim$31\% of the total exposure time of the \cyg\ data set (see Table~\ref{tab:source_sample}). As far as we can tell practically all the remaining data seem to belong to tracks similar to (and usually intermediate between) the ones shown in Figure~\ref{fig:cyg_x-2_sequence} (i.e., none of the other subsets is inconsistent with belonging to such tracks).

Although the tracks in panels K and L do not line up well enough in the HID (contrary to the CD) to be combined into a single track, together they should give a reasonably good idea of the rough shape of a (near-)complete track at this point in the secular progression. It is clear that such a track is similar in shape to those of the Sco-like Z sources.

We note that small secular shifts do occur in some of these tracks; some of the individual segments used have a broad appearance (e.g., the red-colored segment in panel D), which is in some cases probably due to mild secular motion, and in some instances we matched up segments despite their not lining up perfectly in the HID or SID if the overall appearance of the track is only minimally affected by this. There was inevitably sometimes some ambiguity regarding whether certain segments used in a given track rather belonged to a track slightly shifted from the one in question. In particular, the HB upturn was usually observed in short isolated segments in the CD/HID, and it was often hard to judge exactly with which tracks those should be combined. However, the overall conclusions about the secular evolution of the source that we infer from the data are not sensitive to these ambiguities in the combining process. Finally, we note that many of the tracks in Figure~\ref{fig:cyg_x-2_sequence} likely still suffer from some incompleteness, either because data segments that would serve to complete them are simply not available in the \cyg\ data set, or because we felt that there was too much ambiguity in whether candidate segments were appropriate. In particular, we conclude that the track in panel A is likely missing the HB, the tracks in panels B and F are presumably missing (most of) the HB upturn, the one in panel H is missing most of the HB, the track in panel J is missing the NB and HB, and the one in panel K is missing the FB.

\subsubsection{Comparison with \src}\label{sec:comparison_cyg}

\cyg\ exhibits secular evolution that is for the most part very similar to that of \src. As the overall intensity decreases the tracks smoothly evolve in shape from Cyg-like to Sco-like Z tracks. The tracks in \cyg\ panels A--E (Figure~\ref{fig:cyg_x-2_sequence}) are very similar to the Cyg-like Z tracks of \src\ in panels A and B. Taken together, the tracks in \cyg\ panels K and L show that at this point in the secular progression of the source its tracks look similar to (and perhaps intermediate between) those in \src\ panels E and F, the latter of which is very similar to the tracks of the persistent Sco-like Z sources, such as \mbox{GX 17+2} (see, e.g., \citetalias{homan2010}; \citealt{lin2012}). Finally, the tracks in \cyg\ panels M and N look very similar to the tracks in \src\ panels G and H. As the \cyg\ tracks shift and evolve in shape, both the upper and lower vertices quite closely follow straight lines in the CD and HID; as the overall intensity of the tracks decreases, the NB is squeezed between the converging vertex lines and gradually shortens, as observed in \src.

The main differences between \cyg\ and \src\ lie in the fact that \cyg\ is persistently luminous and has never been seen to enter the atoll regime (although the track in panel N does not seem to have an NB, short NB segments seem to exist down to the very lowest intensities observed in the \cyg\ data set) and in the somewhat different FB behavior exhibited by \cyg. In both sources the FB gradually evolves from being a purely dipping FB in the earliest (highest-intensity, Cyg-like) panels to a ``proper'' flaring (Sco-like) FB in later panels. However, in \cyg\ the rotation of the FB is for the most part in the opposite direction to that seen in \src\ in both the CD and HID. In \cyg\ the FB also develops a more complicated morphology in the HID, making a counterclockwise twist of between $180^\circ$ and $270^\circ$ around the lower vertex in panels G and H (such a twist was also observed in \textit{EXOSAT} data by \citealt{kuulkers1996a}) and then assuming a jagged S-like shape in panels I and J that gradually straightens out going from panel I to N. We note that a similar, although less pronounced, S-shaped FB was observed in the HID of panel E in \src.

\subsection{\cir}\label{sec:cir_x-1}

Circinus X-1 (\cir) features some of the richest and most complex phenomenology seen among the known neutron star X-ray binaries, and although it has been extensively studied for over four decades many of its properties remain poorly understood. The binary has an orbital period of $\sim$16.6~days, identified from periodic flaring first observed in the X-ray band \citep{kaluzienski1976}, and later in the radio, infrared, and optical \citep{whelan1977,glass1978,moneti1992}. These flares are thought to be due to enhanced accretion near periastron passage in a highly eccentric orbit \citep[e.g.,][]{murdin1980,tauris1999,jonker2004}. \citet{heinz2013} identify the radio nebula surrounding the binary as a relatively young ($\lesssim$5000~yr old) supernova remnant, which makes \cir\ the youngest known X-ray binary and provides an explanation for the eccentricity of the orbit. This implies that the donor is likely an early-type star \cite[see also][]{jonker2007}. However, type~I X-ray bursts detected from the source \citep{tennant1986a,tennant1986b,linares2010} show that the neutron star is weakly magnetized, which is very unusual for a high-mass companion.

\begin{figure}[t]
\centerline{\includegraphics[width=8.6cm]{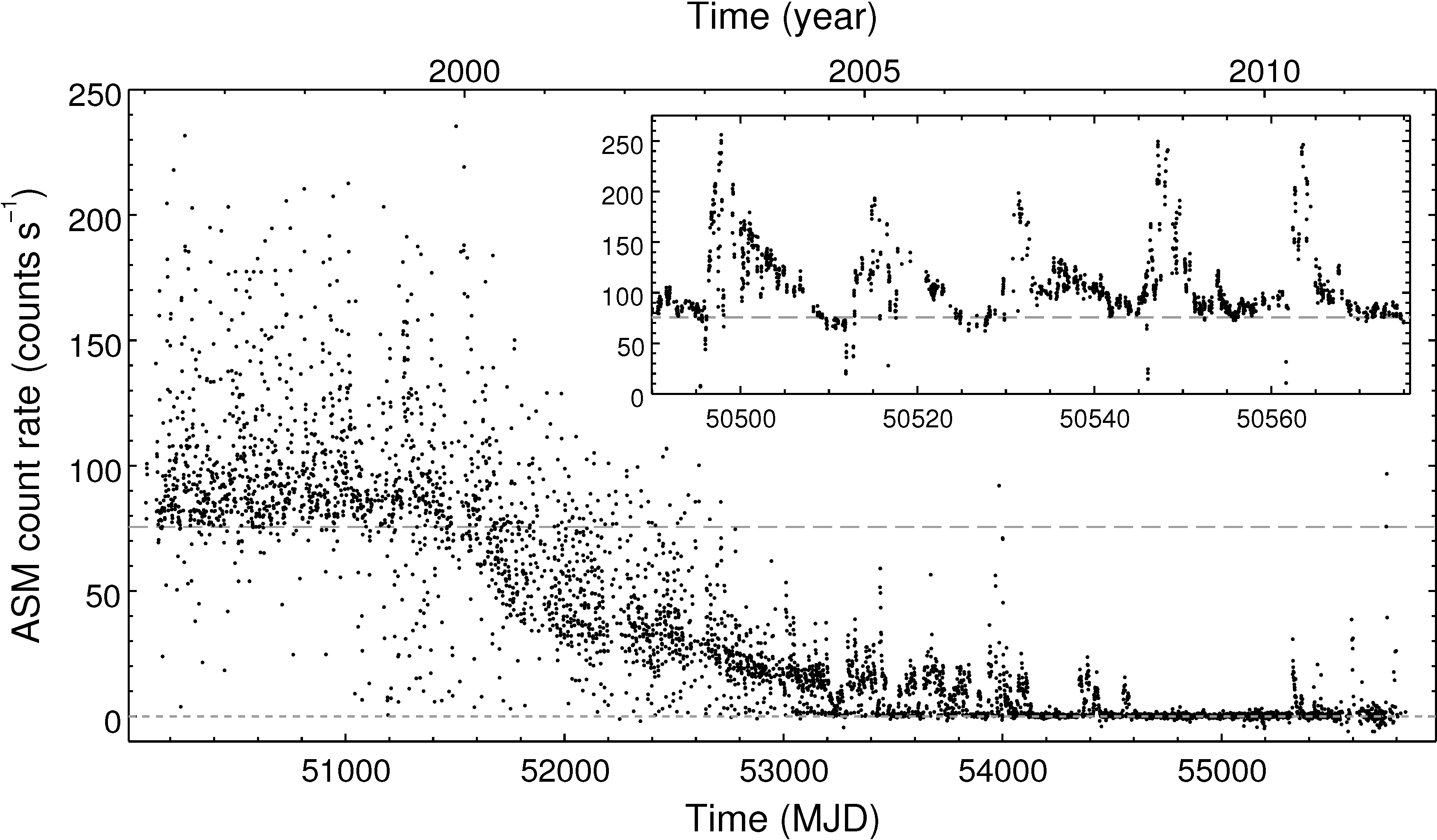}}
\caption{\xte\ ASM light curve of \cir\ covering the entire lifetime of the \xte\ mission. Data points in the main plot are one day averages. The inset zooms in on five orbital cycles in early 1997; each data point there corresponds to a single dwell of the ASM. The long-dashed lines in the main plot and inset show the typical ASM count rate level for the Crab Nebula.}\label{fig:cir_x-1_lc}
\end{figure}

The X-ray emission from \cir\ is highly variable on a wide range of timescales. In Figure~\ref{fig:cir_x-1_lc} we show an \xte\ ASM light curve of the source. The source was in a historically high state during the first few years of the \xte\ mission \citep{saz2003}, with an average flux of \mbox{$\sim$1.3 Crab} (and a maximum observed flux with the ASM of \mbox{$\sim$3.5--4 Crab}). The flux started gradually decreasing in mid-to-late 1999, and kept doing so until the source became undetectable with the ASM and showed no measurable activity over a two-year period in 2008--2010. In 2009 the source was observed with \cxo\ at a flux of only a few tenths of a milliCrab \citep{sell2010}. Since 2010 May, however, the source has shown sporadic activity. When active the source usually exhibits complex variability over the course of an orbital period, featuring both absorption dips and flaring in the X-ray band; the effects of this on the ASM light curve can easily be seen in Figure~\ref{fig:cir_x-1_lc}. In the inset we show five orbital cycles from the bright phase of the source early in the \xte\ mission. X-ray flaring is strongest during the first few days after it commences; it then gradually decreases in strength as the orbit progresses. The source usually exhibits strong X-ray dipping during the last $\sim$0.5$-$1~days before the onset of flaring and then intermittent dips for up to two days afterwards \citep{shirey1998}; we discuss this dipping in more detail below.

\citet{oosterbroek1995} reported indications of both Z and atoll behavior in \textit{EXOSAT} data of \cir. \citet{shirey1999a} analyzed data from 10~days of \xte\, PCA observations in 1997. They detected all three Z branches---although with some differences in shape compared to the classic Z sources---and found that the tracks moved around the CD/HID and evolved in shape. The identifications of the Z branches were supported by timing data.

\subsubsection{Analysis}\label{sec:cir_x-1_analysis}

\begin{figure}[t]
\centerline{\includegraphics[width=8.6cm]{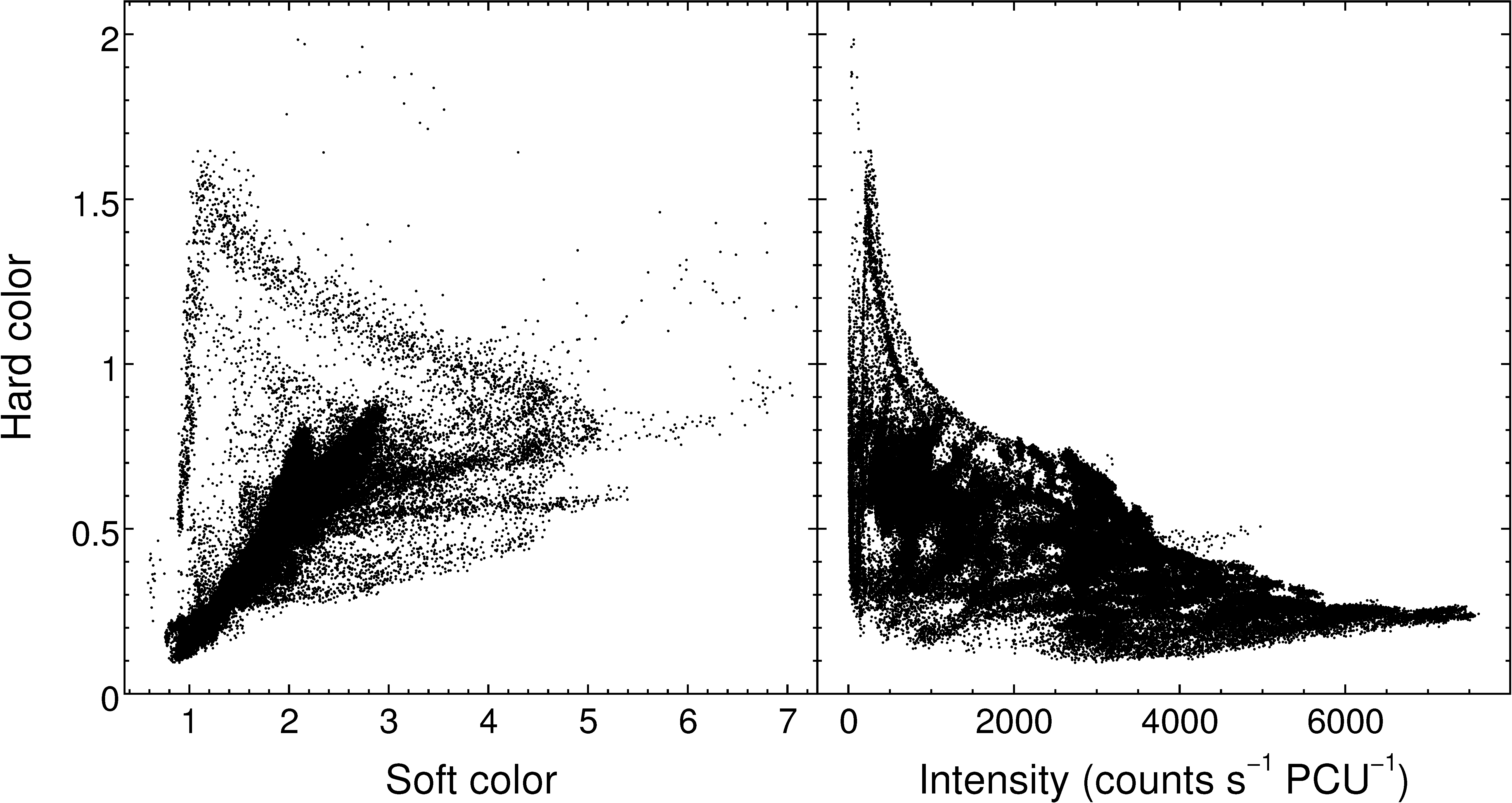}}
\vspace{0.2cm}
\centerline{\includegraphics[width=8.6cm]{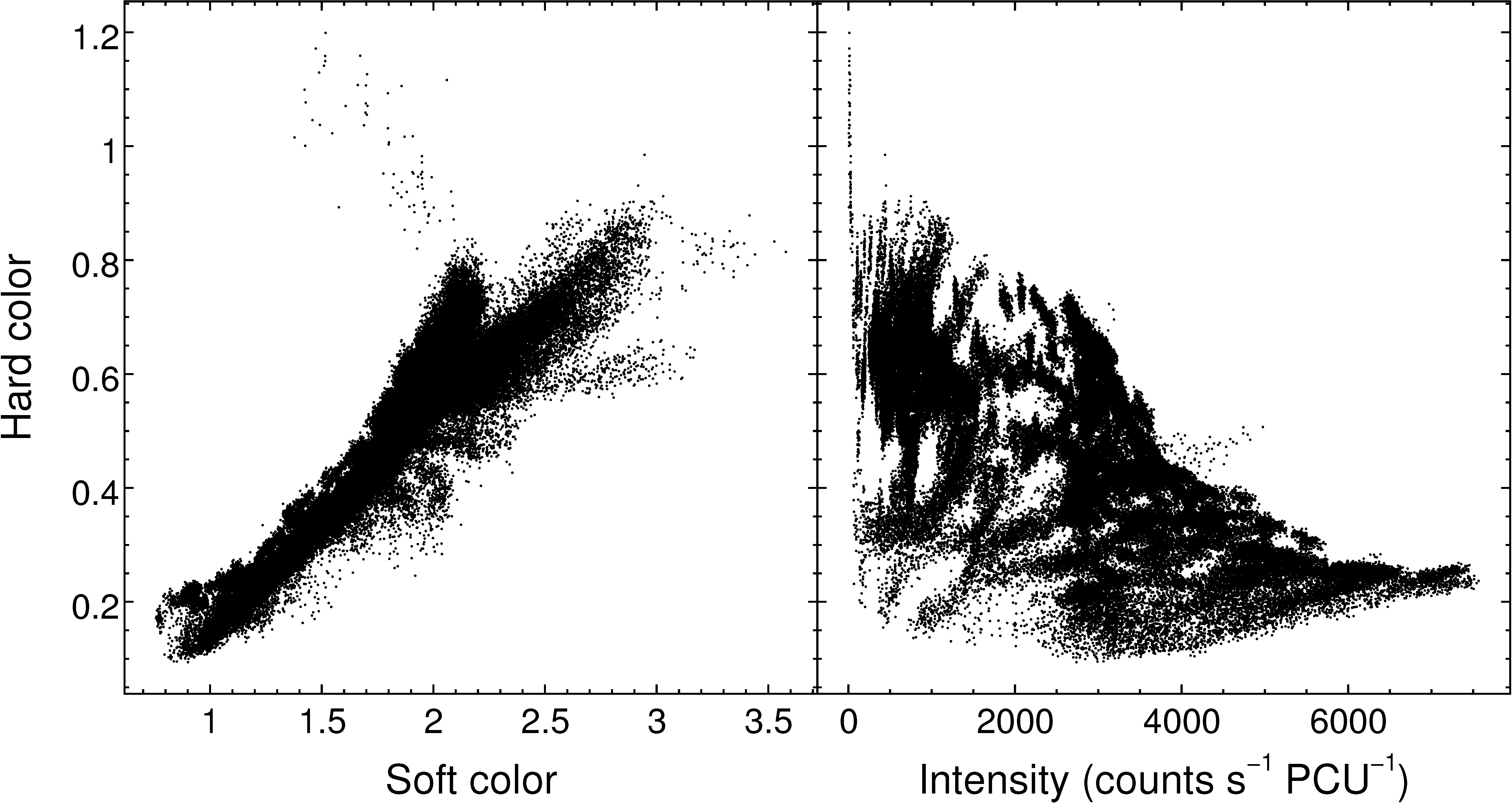}}
\caption{CD and HID representing the entire \xte\ PCA data set of \cir\ before (upper plot) and after (lower plot) the removal of data affected by absorption. A small portion of the data at very high soft and hard color values falls outside the diagrams in the upper panel. Note the differences in scale in soft and hard color between the upper and lower panels.}
\label{fig:cir_x-1_all_data}
\end{figure}

\begin{figure*}[t]
\centerline{
\includegraphics[width=5.71cm]{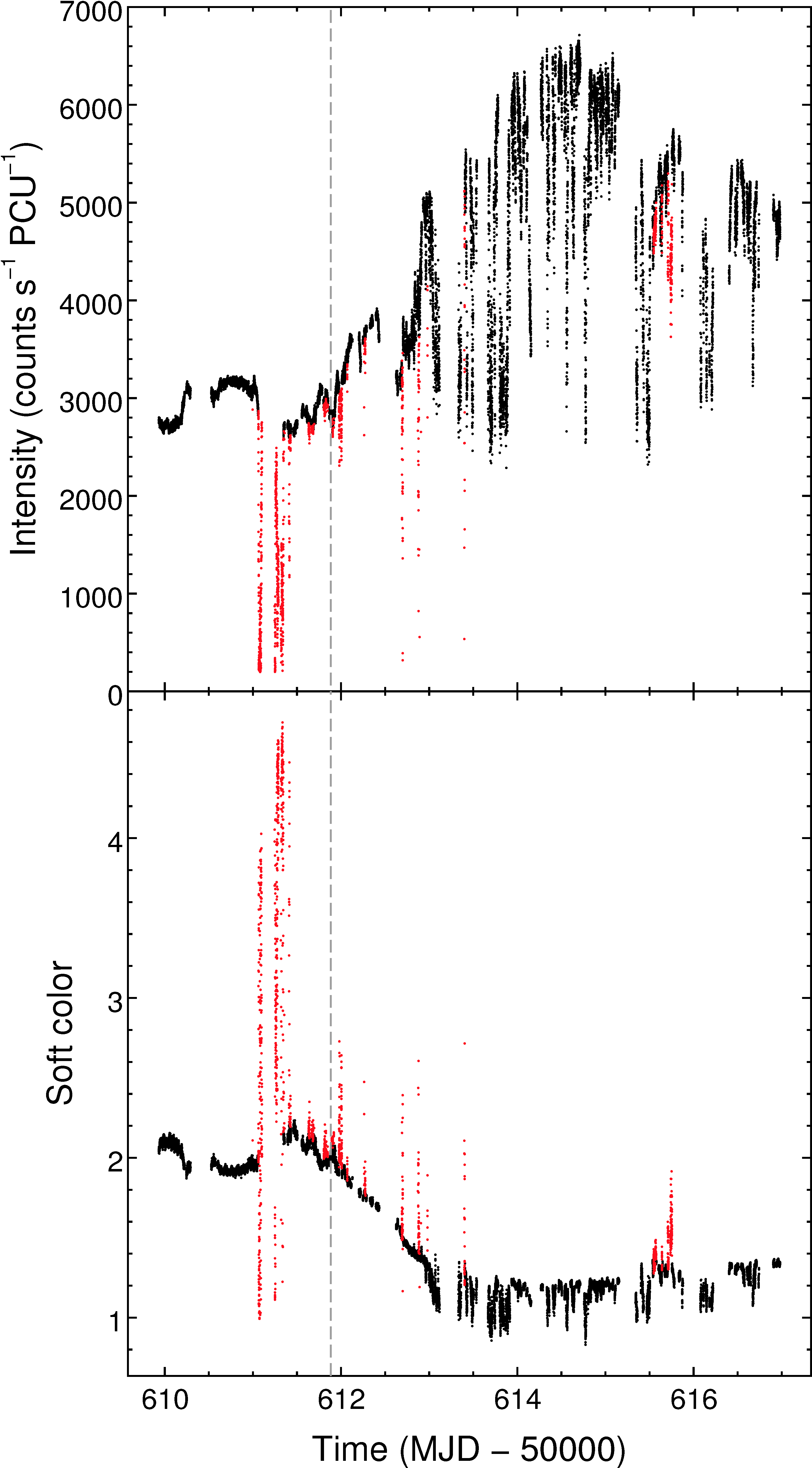}
\hspace{0.5cm}
\includegraphics[width=10.5cm,trim=0 -280 0 0]{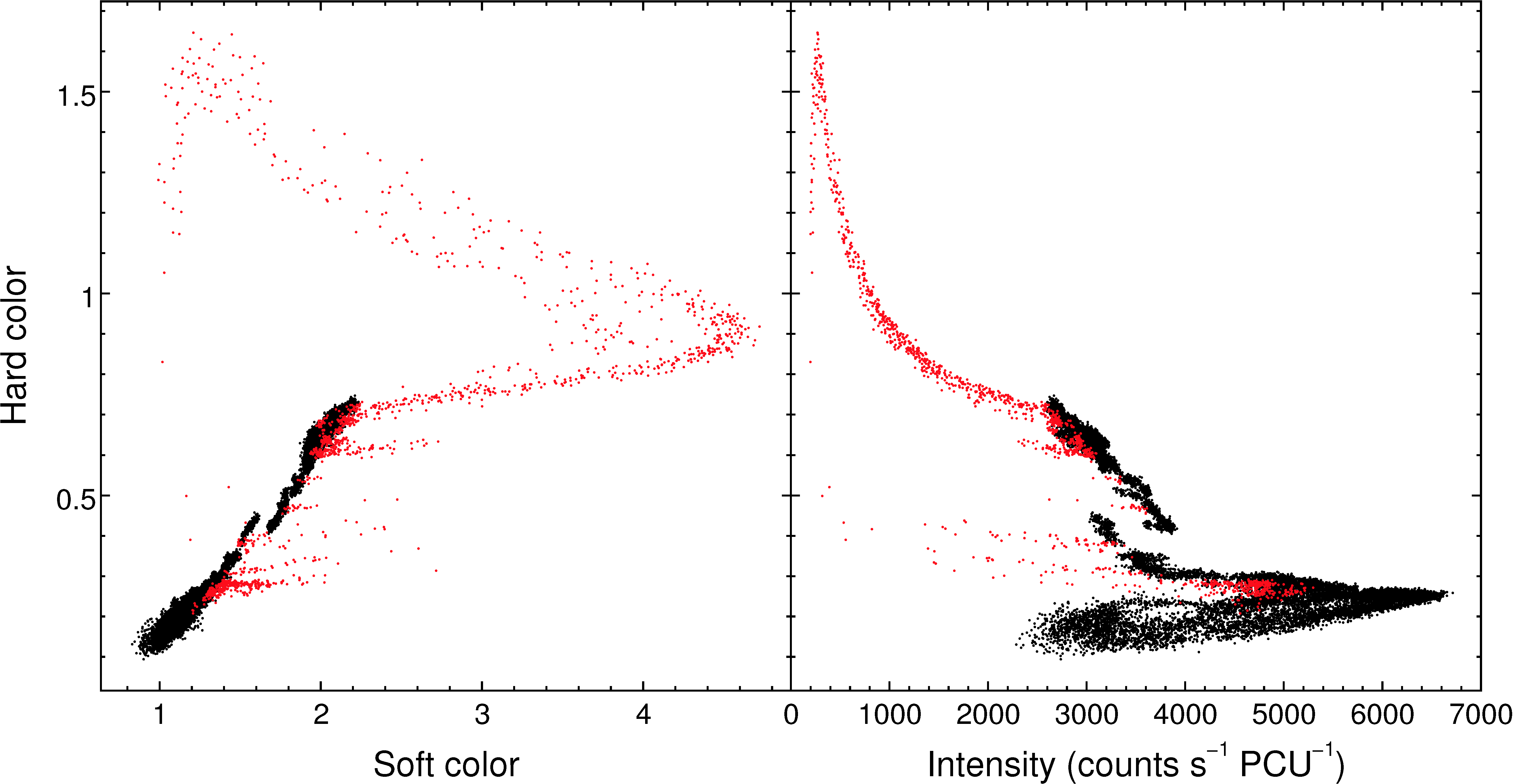}
}
\caption{Diagrams demonstrating the removal of data affected by absorption from 7~days of observations of \cir\ in 1997 June; the left plot shows intensity and soft color as a function of time, and the right plot shows a CD/HID. We identified the red data points with absorption and removed them from the data set before proceeding with further analysis. In the left plot the dashed vertical line shows the time of zero phase according to the dipping ephemeris of \citet{clarkson2004}.}\label{fig:cir_x-1_dipping_removal}
\end{figure*}

In our analysis we considered all 811 individual ObsIDs for pointed observations of \cir\ made during the \xte\ mission (18 of which contained no useful data). A CD/HID using the entire data set (which spans $\sim$15 years) is shown in the upper plot in Figure~\ref{fig:cir_x-1_all_data}. The diagrams are heavily affected by both absorption and secular shifts and shape changes in the source tracks. As mentioned above, most of the dipping occurs close to the time of presumed periastron passage, which is often associated with flaring. \citet{clarkson2004} fitted a quadratic ephemeris to the times of dips observed in \xte\ ASM data from 1996 to 2003 and found that it provides a good predictor of the X-ray light curve. The strongest dipping often produces a characteristic track in the CD and HID; this track goes up to high color values and has two sharp bends in the CD. This can be seen in Figures~\ref{fig:cir_x-1_all_data} and \ref{fig:cir_x-1_dipping_removal}; in the latter we show various diagrams for 7~days of observations in 1997 June, which we discuss further below. (The observations used in Figure~\ref{fig:cir_x-1_dipping_removal} constitute the bulk of the data analyzed by \citealt{shirey1999a}). The shape of these dipping tracks can be understood if the X-ray emission is composed of two components: a bright component that is subject to heavy absorption and a faint component that is unaffected by the absorption, the latter perhaps due to X-rays from the central source scattered into our line of sight by surrounding material \citep{shirey1999b}. We note that the source also often exhibits shallower dips (also seen in Figure~\ref{fig:cir_x-1_dipping_removal}) which do not result in a (full) track of the sort described above.

\begin{figure*}[t]
\centering
\includegraphics[height=11.9cm]{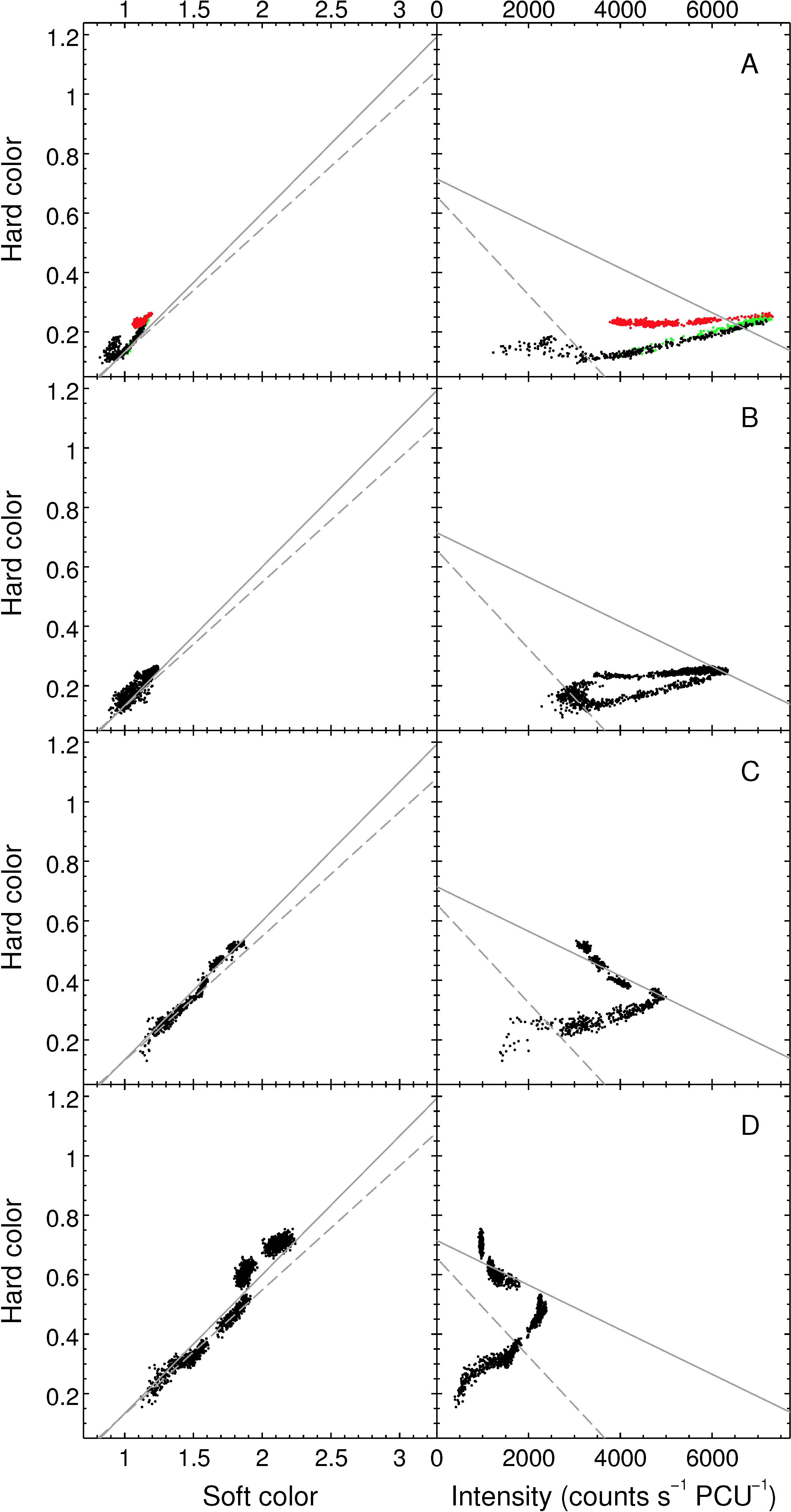}
\hspace{-0.04cm}
\includegraphics[height=11.9cm]{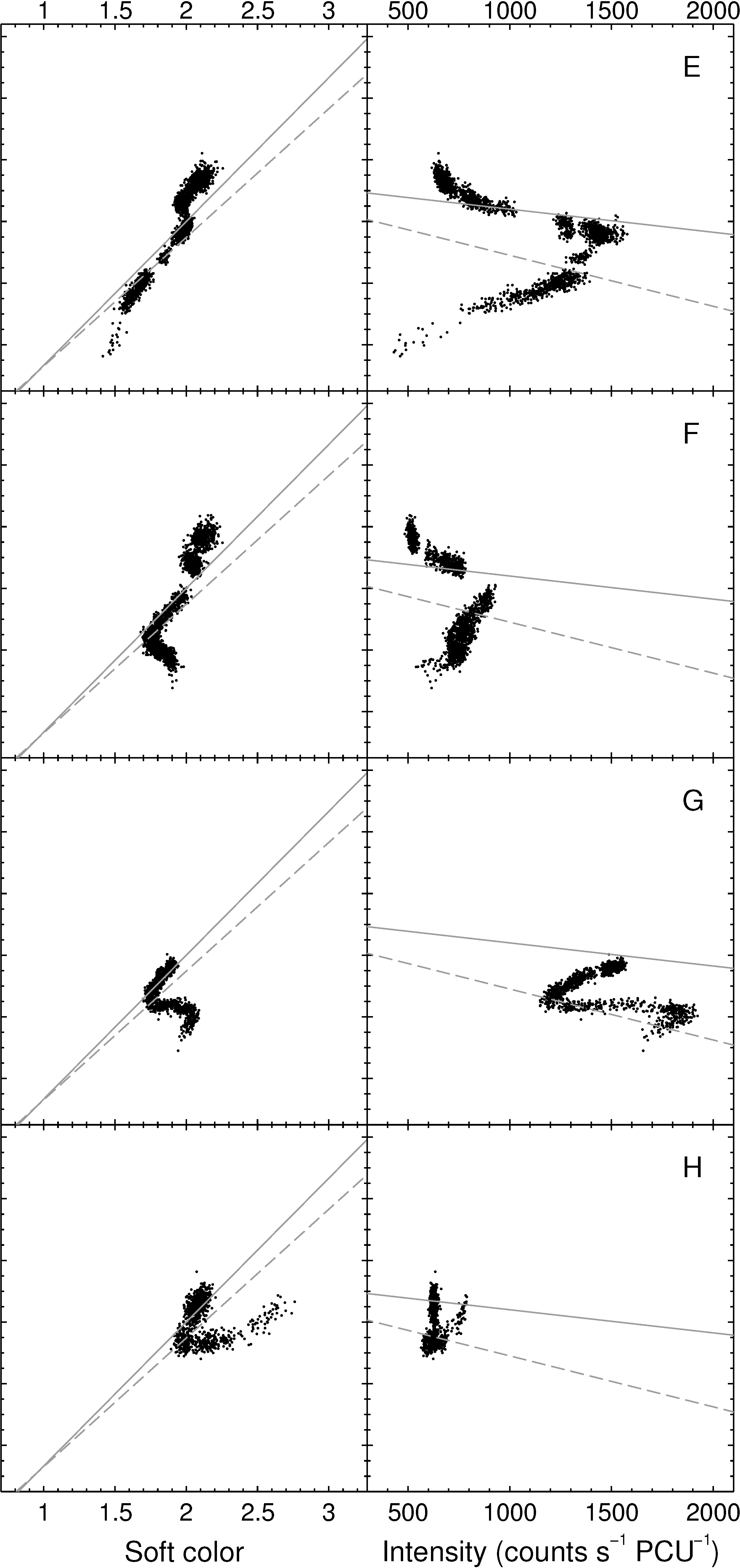}
\hspace{-0.078cm}
\includegraphics[height=11.9cm]{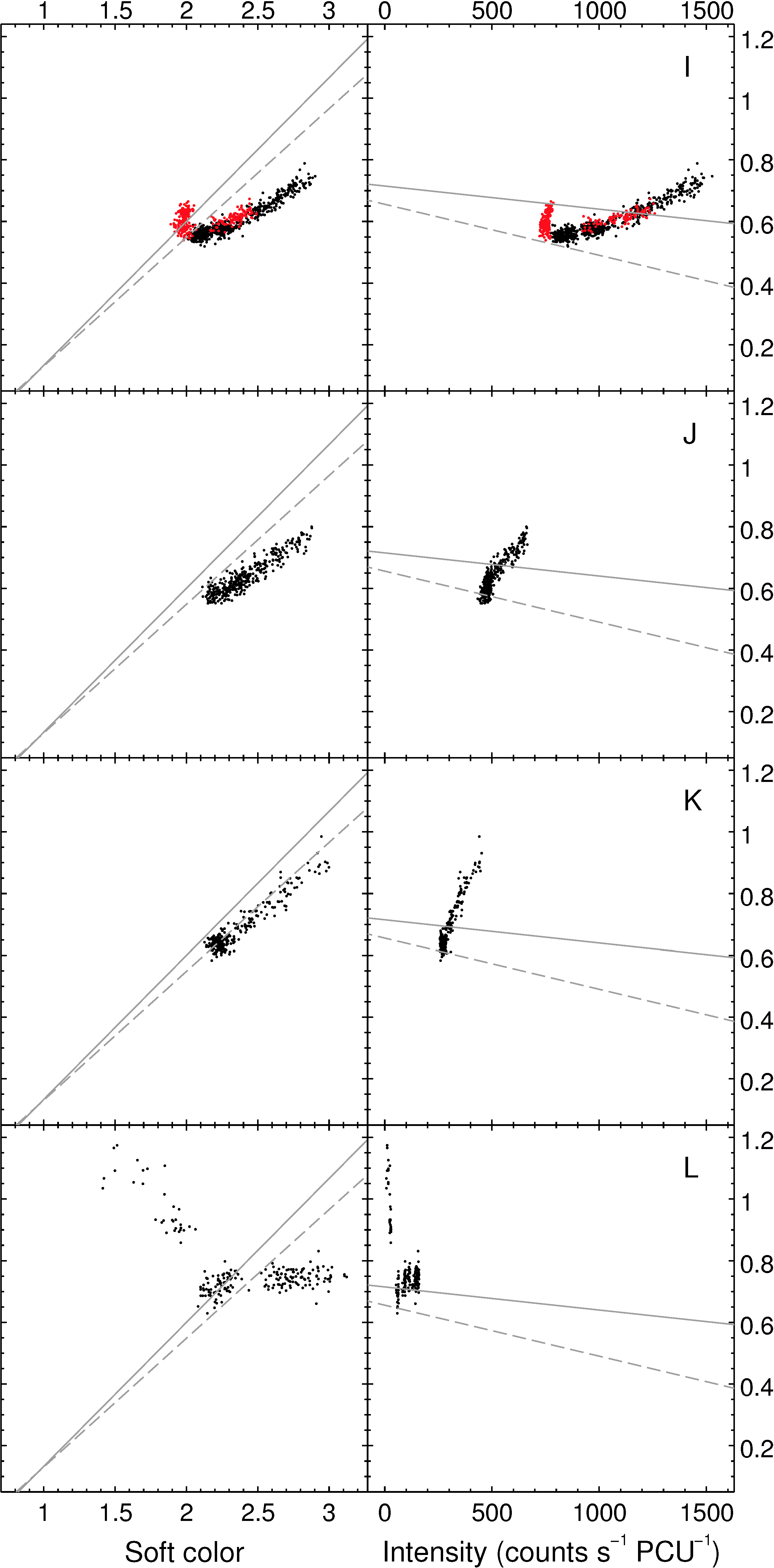}

\caption{A sequence of CDs/HIDs for \cir\ illustrating the secular evolution of the source. The data points in each panel were obtained within a relatively short period of time, ranging from $\sim$4~hr to $\sim$7~days, except for panels A and I, which combine three and two segments, respectively, widely separated in time (with each segment shown in a particular color). The dashed and solid lines are lower and upper vertex lines similar to those shown for \src\ and \cyg. Data are binned to a minimum of 16,000~counts per data point, except for IS and EIS points in panel L, which are binned to a minimum of 32,000~counts. Note the change in the intensity scale between the different HID columns.}\label{fig:cir_x-1_sequence}
\end{figure*}

Before proceeding with our analysis of the \cir\ data set we removed by eye (to the extent possible) data points affected by absorption dips. An example of this is shown in Figure~\ref{fig:cir_x-1_dipping_removal}, where removed data points are colored red. As can be seen in the figure, during these observations the source showed intense dipping shortly before the start of flaring (at day $\sim$611.5), producing the characteristic dipping track in the CD/HID. Some shallower dips were then observed during the first $\sim$2~days of flaring. We note that the periodic ``flaring'' during an orbital cycle is in general associated with motion along all three (Cyg-like) Z branches, rather than being exclusively associated with motion along the FB (which is mostly a dipping FB in these particular observations). As is apparent from the figure, it is almost impossible to identify absorption dips on the basis of the light curve alone during periods of flaring. However, tracking the behavior of the soft color (and, to a lesser extent, the hard color) as a function of time, as well as inspecting the CD and HID, greatly aids in identifying dipping. In addition, we took into account the dipping ephemeris of \citet{clarkson2004} when performing the removal, since the vast majority of dipping events take place in the $\sim$1--1.5~days immediately before or after phase~0. However, we note that (shallow) dips are sometimes seen later in an orbital cycle; one example can be seen in Figure~\ref{fig:cir_x-1_dipping_removal}. In general, when unsure whether a given data segment was afflicted by dipping or not, we in general opted to rather err on the side of caution and remove the segment in question. The ``cleaned'' CD/HID resulting from our manual removal of absorption-affected data is shown in the lower plot in Figure~\ref{fig:cir_x-1_all_data}. It was of course unavoidable that some instances of minor dipping remain and that a small amount of unaffected data be removed. However, we expect that the effects of this on the conclusions we draw from the data are negligible.

After removing data points affected by absorption dips, as described above, we organized the data into subsets, similar to our \cyg\ analysis described in the previous section. This resulted in $\sim$300 subsets spanning periods as long as several days, although most are less than a day in length. From these we created a sequence of 12 CD/HID (partial) tracks, shown in Figure~\ref{fig:cir_x-1_sequence}. These tracks were chosen to illustrate as best possible the overall secular behavior of the source. In all cases except two the data in a given panel are from a single subset, whose time intervals range from $\sim$2~hr to $\sim$7~days. The exceptions are panels A and I, where 2--3 independent segments (shown in different colors) were combined. It was not viable to form a whole sequence of combined and more complete tracks, as we did for \cyg, since the data for \cir\ show a much greater dynamic range and do not sample the HID densely enough for this. Overall, the upper and lower vertices of the \cir\ tracks move systematically up and to the right in the CD, and up and to the left in the HID, as the tracks evolve in shape---similar to the behavior seen in \src\ and \cyg. The paths of the vertices can be approximated by the straight lines shown in Figure~\ref{fig:cir_x-1_sequence}. We order the tracks based on the position of the lower and/or upper vertex in the CD. (When both are present, the two vertices give consistent results.) In cases where some ambiguity remains due to very similar vertex locations (in particular, panels F and G), we use the shape of the track in the CD to decide the ordering, in particular the FB (i.e., we choose the ordering that produces a gradual evolution in the shape of the track in the CD). The combined exposure time of the data shown in Figure~\ref{fig:cir_x-1_sequence} is $\sim$10\% of the total exposure time of the \cir\ data set. The data subsets (time intervals) and ObsIDs used in each panel are listed in Table~\ref{tab:cir_x-1_obs}.

We note that a few of the tracks in Figure~\ref{fig:cir_x-1_sequence} show some signs of secular motion in the HID (but much less in the CD); however, in most cases this has a negligible effect on the overall appearance of the track. The instance where this is most noticable is in the HID of panel E, where the upper parts of the track (NB and especially HB) show shifting toward lower intensities relative to the FB. A similar, but smaller, shift affects the HID in panel F. We also note that the behavior of \cir\ is not as regular as that of \src\ and \cyg, and the vertices in the HID (and to a lesser extent in the CD) often deviate significantly from the lines shown. As will be discussed in Section~\ref{sec:discussion}, the relationship between a track's shape and its location in the HID is also not nearly as tight for \cir\ as for \src\ and \cyg, where the motion of both vertices in both the CD and HID, as well as the decrease in overall intensity, is monotonic along the entire sequence. The behavior of \cir\ is more regular in the CD than the HID, which is why we chose to use the CD as the basis for our ordering of the tracks. However, apart from some irregularities in panels F--I, the \cir\ sequence does show a gradual decrease in overall intensity.

Panel L shows what seems to be an atoll transition from the banana branch through the IS and to an EIS. These data were obtained in 2010 May/June as the intensity steadily decreased over a $\sim$7~day period (part of a $\sim$50~day minioutburst following two years of nondetection by the \xte\ ASM). Observations preceding this 7~day period show what looks like a sequence of segments at successively lower count rates from tracks similar to the one in panel K. Figure~\ref{fig:cir_x-1_atoll} shows the same data as in panel L (in black) but with more binning. \citet{dai2012} studied the spectral evolution of the source during the entire 2010 May--June minioutburst with spectral fitting and also concluded that the source transitioned from the atoll soft to hard state. In 2010 August the source showed a similar transition (again with an intensity decline over $\sim$7~days), which we show with red data points in Figure~\ref{fig:cir_x-1_atoll}. We note that the apparent decrease in hard color with decreasing intensity in the EIS in both cases may well be due to soft diffuse background emission (which could not be subtracted) affecting the data points at the lowest intensities.

\subsubsection{Comparison with \src\ and \cyg}\label{sec:comparison_cir}

Like \src, \cir\ has been observed in all NS-LMXB subclasses (Cyg-like Z, Sco-like Z, atoll). The overall secular evolution has many similarities to that observed for \src\ and \cyg, and many of the individual tracks have shapes similar to those seen for those two sources. Progressing along the \cir\ sequence the NB grows shorter and rotates counterclockwise in the HID, while the HB rotates clockwise as it shortens, similar to what was seen for \src\ and \cyg. As in those two sources the FB shows the most complex behavior of the three branches in \cir. In the CD of \cir\ the FB gradually rotates counterclockwise and evolves in shape in a similar fashion to \src. In the HID the behavior of the FB is more irregular in \cir\ than \src\ and \cyg, but overall it seems to rotate counterclockwise as it evolves from a dipping FB in the higher-intensity tracks to a Sco-like FB at lower intensities.

In panel G, the \cir\ FB in both the CD and HID has a shape similar to those of the persistent Cyg-like Z sources \mbox{GX 340+0} and \mbox{GX 5--1} (see, e.g., \citealt{jonker1998}; \citealt{jonker2002}; \citetalias{homan2010}). We note that in the CD of \src\ the shape and orientation of the FB in panels A and B seems intermediate between \cir\ panels E and F, whereas the \src\ FB in panels C (which has a very incomplete track) and D seems intermediate between \cir\ panels G and H. \src\ may therefore have traced out a CD (and HID) track similar to those of \mbox{GX 340+0} and \mbox{GX 5--1} in between panels B and C; this portion of the secular progression of the source was missed due to a gap in \xte\ coverage and rapid secular evolution. However, we note that in none of the four sources compared in this paper do we see a full track in both the CD and HID where the shapes of all three spectral branches closely match those of \mbox{GX 5--1} and \mbox{GX 340+0}.

\begin{figure}[]
\centerline{\includegraphics[width=8.6cm]{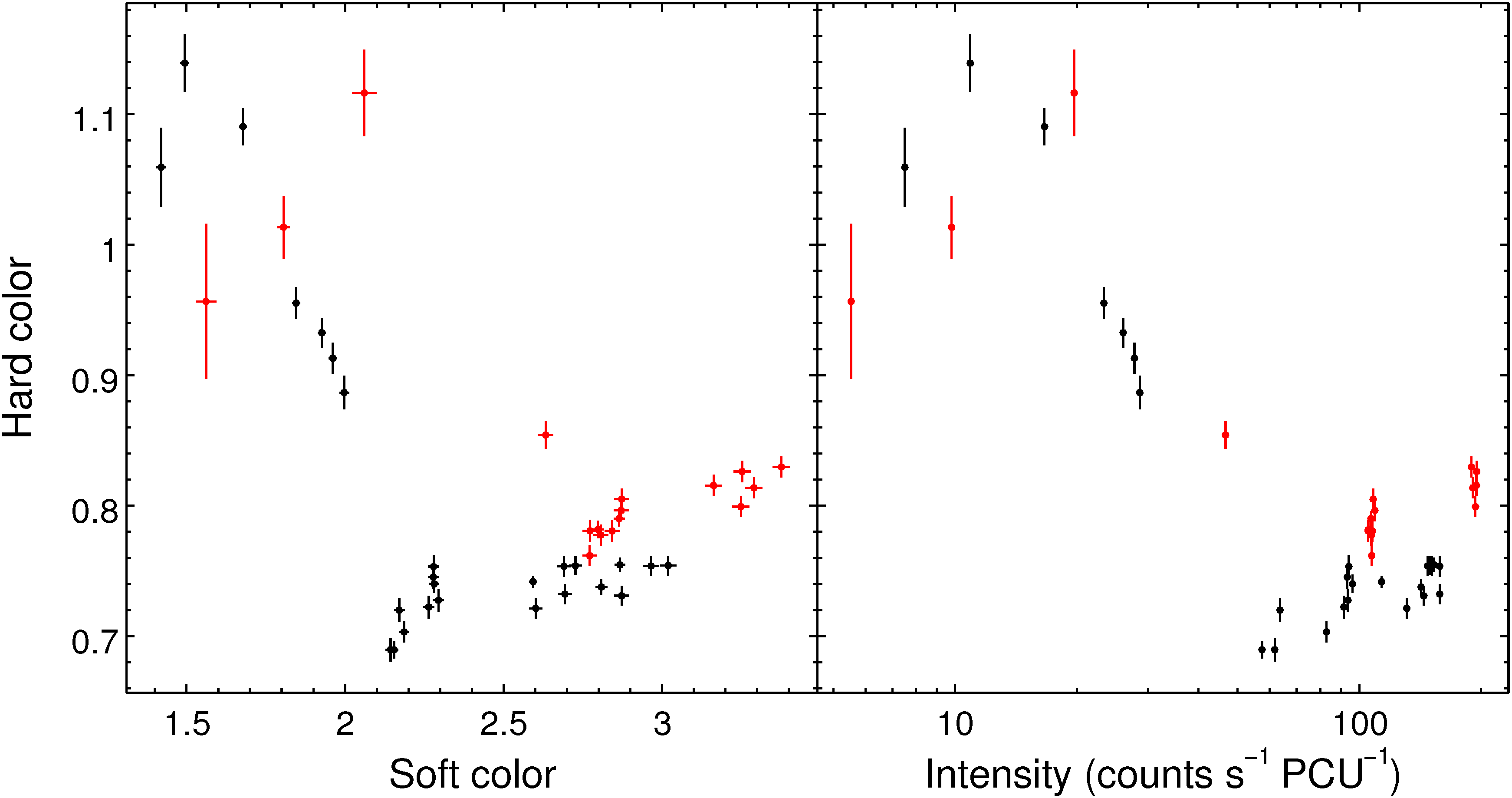}}
\caption{CD/HID showing two instances of \cir\ undergoing an atoll transition from the soft to hard state. Black data points are from a $\sim$7~day period in 2010 May/June (also shown in panel L in Figure~\ref{fig:cir_x-1_sequence}) and red data points are from a $\sim$7~day period in 2010 August. The data were binned with a minimum of 128,000~counts per data point. In some cases entire observations did not have that many counts and were combined with other ones close in time.}\label{fig:cir_x-1_atoll}
\end{figure}

The atoll transitions of \cir---especially the one in 2010 May/June---resemble the one of \src\ (see Figures~\ref{fig:1701_atoll}~and~\ref{fig:cir_x-1_atoll}); one notable similarity is that the soft color decreases significantly throughout the transitions, in contrast to the increase observed in most atoll sources \citep[e.g.,][]{mythesis2011}. However, there is also a striking difference between the atoll transitions of the two sources: the data from the banana branch preceding the ascent to the hard state extend to very high soft color for \cir, occupying parameter space in the CD never explored before by the source in the \xte\ archive except during absorption dips or (partly) during traversals to the tip of the (nondipping) FB. The data points in question are unlikely to be associated with the FB, given the small changes in intensity and hard color observed, and the long timescale involved (more than a day). This pre-atoll-transition behavior of \cir\ is in stark contrast to \src, where no such excursion to high soft color values was seen, and the region occupied by the source in the CD in selection L before moving to the IS was a logical extension of the movement of the source in the preceding selections. We also note that the observations in 2010 May/June were the first time in almost 15~years of \xte\ observations that \cir\ was seen to transition to the atoll hard state.

\setlength\bigstrutjot{4pt}
\tabletypesize{\footnotesize}

\begin{deluxetable*}{clclc}[]
\tablewidth{18.0cm}
\tablecolumns{5}
\tablecaption{Time Intervals and Observations Used to Create \cir\ Tracks \label{tab:cir_x-1_obs}}
\tablehead{ & & \colhead{Interval Length} & & \colhead{Exp.\ Time} \\[0.3ex]
\colhead{\hspace{-0.13cm}Panel} & \colhead{\hspace{-0.65cm}Time Interval (MJD)} & \colhead{days (hr)\tablenotemark{a}} & \colhead{\hspace{-4.4cm}ObsIDs} & \colhead{ks (hr)\tablenotemark{b}}}
\startdata
A				& 50365.216--50365.307		& 0.09 (2.2)\phn\phn		& 10068-08-02-00							& 5.5 (1.5) \\[0.6ex]
				& 50497.364--50497.437	(G)	& 0.07 (1.8)\phn\phn		& 20095-01-01-00 							& 4.8 (1.3) \\[0.6ex]
				& 50711.692--50711.763	(R)	& 0.07 (1.7)\phn\phn		& 20095-01-18-00 							& 5.9 (1.6) \bigstrut[b] \\
\hline
B				& 50613.800--50614.155		& 0.36 (8.5)\phn\phn		& 20094-01-02-[04,040]\tablenotemark{c}			& 21.6 (6.0)\phn \bigstrut \\
\hline
C				& 51603.985--51604.316		& 0.33 (8.0)\phn\phn		& 40059-01-01-[00,02]\tablenotemark{c}			& 13.1 (3.6)\phn \bigstrut \\
\hline
D				& 52615.803--52618.276		& 2.47 (59.4)\phn		& 70020-01-[02-(01:02),03-01,03-04,04-(00:04)]	& 37.9 (10.5) \bigstrut \\
\hline
E				& 53013.822--53017.845		& 4.02 (96.6)\phn		& 70020-03-04-[00:20]						& 34.6 (9.6)\phn \bigstrut \\
\hline
F				& 53163.821--53165.483		& 1.66 (39.9)\phn		& 80027-02-[02-02,02-06,03-(00:02),03-06]		& 31.0 (8.6)\phn \bigstrut \\
\hline
G				& 51831.981--51832.459		& 0.48 (11.5)\phn		& 50136-01-04-[00:06]						& 18.7 (5.2)\phn \bigstrut \\
\hline
H				& 53168.550--53168.789		& 0.24 (5.7)\phn\phn		& 80027-02-04-01							& 14.1 (3.9)\phn \bigstrut \\
\hline
I				& 52787.744--52787.894	(R)	& 0.15 (3.6)\phn\phn		& 80114-04-01-[02:04] 						& 4.4 (1.2) \bigstrut[t] \\[0.6ex]
				& 52951.992--52952.229		& 0.24 (5.7)\phn\phn		& 80027-01-01-[02:03]						& 9.4 (2.6) \bigstrut[b] \\
\hline
J				& 53003.507--53003.590		& 0.08 (2.0)\phn\phn		& 70020-03-01-01							& 5.6 (1.6) \bigstrut \\
\hline
K				& 53271.381--53271.787		& 0.41 (9.8)\phn\phn		& 90025-01-02-[02,24,25,27]					& 7.9 (2.2) \bigstrut \\
\hline
L				& 55343.361--55350.716		& 7.36 (176.5)			& 95422-01-[03-(03:04),04-(00:13)]				& 49.4 (13.7) \bigstrut
\enddata
\tablenotetext{a}{The interval length is shown in units of both days and hours.}
\tablenotetext{b}{The total exposure time is shown in units of both ks and hours.}
\tablenotetext{c}{The early parts of observations 20094-01-02-040 and 40059-01-01-00  were omitted due to secular motion.}
\tablecomments{Subsets colored in red/green in Figure~\ref{fig:cyg_x-2_sequence} are denoted by (R)/(G) in the Time Interval column. In the ObsIDs column a colon denotes a range.}
\end{deluxetable*}

The \cir\ track in panel D bears a strong resemblance in shape to the highest-intensity tracks (panels A and B) of \src\ (in particular when the HID is plotted on the same scale) and occupies a similar color range (especially in hard color). But \cir\ also shows tracks that reach much higher intensities (panels A--C). Going from panel D to A in the \cir\ sequence, the overall color values of tracks decrease and the color range spanned drops sharply as the intensity swings become larger, resulting in very stretched and flat tracks in the HID. In these ``extreme'' Cyg-like Z tracks in panels A and B \cir\ exhibits much lower color values than ever observed for any of the other three sources analyzed in this paper, and it is striking how small the color variations along the tracks are despite the large changes in intensity.

\vspace{0.7cm}

\subsection{\gx}\label{sec:gx_13+1}

\gx\ is a bright X-ray binary whose classification as a Z or atoll source is ambiguous, although it has usually been labeled an atoll source. \citet{hasinger1989} classified \gx\ as a bright atoll source based on CDs and power spectra from \textit{EXOSAT} observations. In \xte\ observations from 1996, \citet{homan1998} discovered a \mbox{57--69 Hz} QPO, which showed similar behavior to the horizontal-branch oscillation (HBO) seen in Z sources. \citet{schnerr2003} performed a combined CD, HID, and power-spectral analysis of a large number of \xte\ observations made in 1998. They found that the source traced out, on a timescale of hours, a curved two-branched track in the CD, which showed strong secular motion on a timescale of $\sim$1~week. The shape of the track was similar to the lower part (IS, LB, UB) of an atoll track or the NB/FB part of a Z track; the location of the vertex between the two branches was seen to approximately follow a straight line in the CD. They also found that the source showed peculiar CD/HID and rapid-variability behavior compared to most other Z or atoll sources, but overall they favored an atoll classification. \citet{homan2004} analyzed two simultaneous \xte/radio observations of \gx\ performed in 1999. Based on the results of spectral fits, rapid-variability properties, behavior in the radio band, and the scarcity of type~I X-ray bursts observed from the source since its discovery, they concluded that the properties of \gx\ were more similar to Z sources than atolls.

\subsubsection{Analysis}\label{sec:gx_13+1_analysis}

\begin{figure}[t]
\centerline{\includegraphics[width=8.6cm]{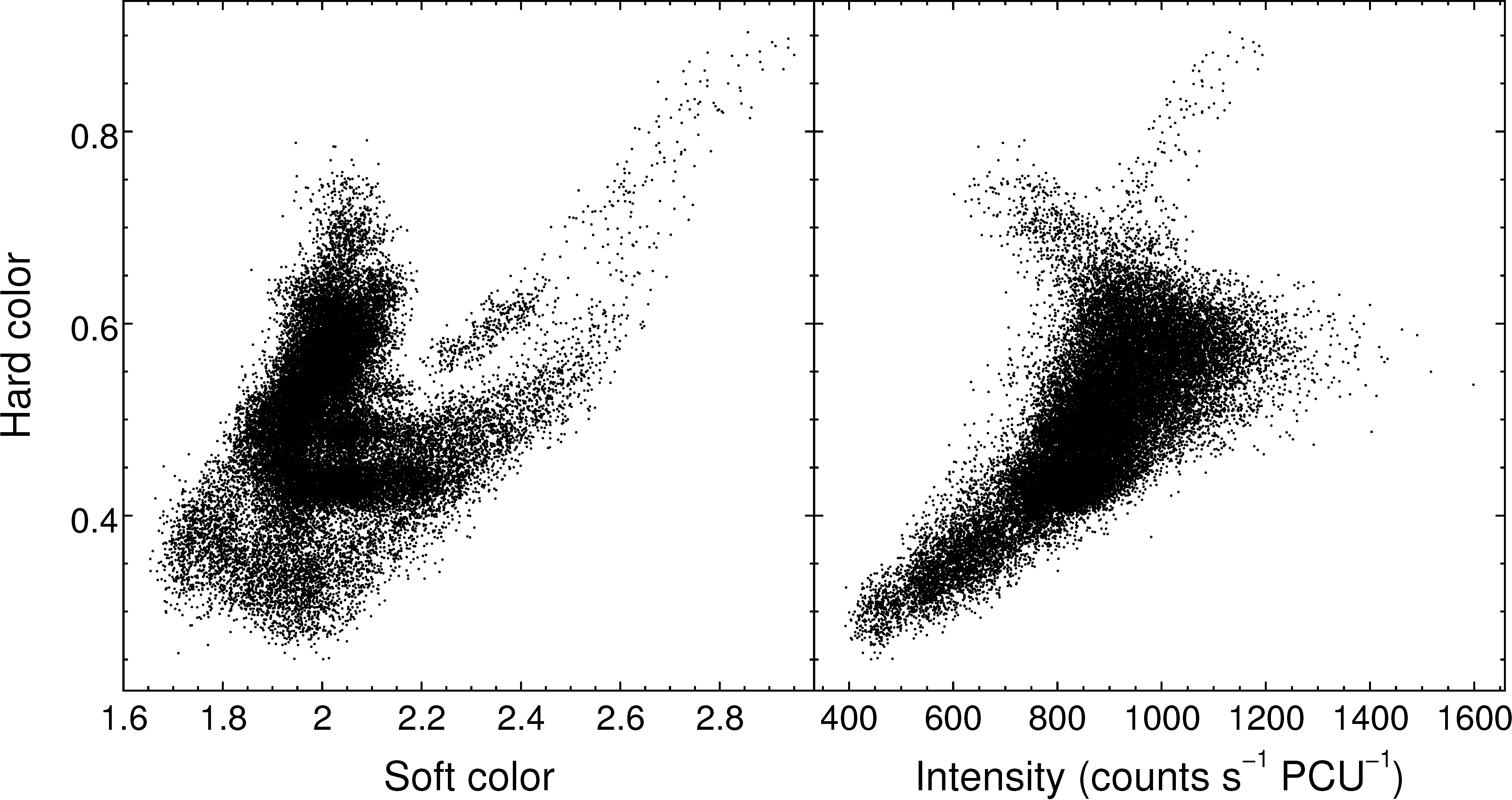}}
\caption{CD and HID representing the entire \xte\ PCA data set of \gx.}\label{fig:gx_13+1_all_data}
\end{figure}

In our analysis we considered all 92 individual ObsIDs for pointed observations of \gx. These span a period of $\sim$14~years. In Figure~\ref{fig:gx_13+1_all_data} we show a CD/HID based on all the \gx\ data; strong secular motion is apparent. However, we note that \gx\ overall shows the smallest range in secular evolution among the four sources studied here (e.g., as quantified by the range in soft or hard color over which the lower vertex is observed), and the other three sources show secular motion on shorter timescales than does \gx. As for \cyg\ and \cir\ we organized the data into subsets; this resulted in $\sim$50 such sets, which could in a few cases span intervals as long as several days with little or no visible secular motion. We constructed a sequence of six CD/HID tracks (shown in Figure~\ref{fig:gx_13+1_sequence}), which illustrates the secular evolution of the source. Similar to our analysis of \cyg\ we combined subsets widely separated in time in three of these tracks to create more complete tracks than otherwise available and thereby give a fuller depiction of the overall secular behavior of the source. For these three tracks we also indicate in red a representative subset obtained in a short time interval. The number of individual subsets used for each track ranges from 1 to 4; these are listed in Table~\ref{tab:gx_13+1_obs} along with the corresponding ObsIDs. Similar to our \cyg\ and \cir\ analysis we order the tracks based on the location of the lower vertex in the CD, starting at the lowest soft and hard color. (See discussion of this vertex in the following paragraph.) The data used in the six tracks together constitute $\sim$42\% of the total exposure time of the \gx\ data set. We note that in a few of the subsets used we do see indications of small secular shifts---especially in panels B and F in Figure~\ref{fig:gx_13+1_sequence}. These shifts are apparent mostly in the HID (and SID), rather than the CD. However, the overall conclusions we draw from these tracks and the sequence as a whole are not affected by this.

As can be seen from Figure~\ref{fig:gx_13+1_sequence} the tracks in the CD mostly have a two-branched form. The vertex between these branches follows rather closely a straight line (as observed by \citealt{schnerr2003}), which we show in the figure. We classify \gx\ as a Z source based on its overall secular evolution in the CD/HID, the timescales on which it traces out its CD/HID tracks, and its rapid-variability properties. (We discuss this further in Section~\ref{sec:gx_13+1_discussion}.) We identify the vertex in the CD as the lower Z track vertex and the branches to the left/right of the vertex line in the CD as the normal/flaring branch. In addition, the tracks in panels D and E include subsets that we identify as the HB (plus upturn). In the CD these segments look very similar to excursions onto the HB and HB upturn in the Cyg-like Z tracks of \src, \cyg, and \cir. In the HID these segments also stand out and are located above and to the left of the rest of the track. We note that there is considerable ambiguity regarding where within panels C--E to place these presumed HB segments. What we show in Figure~\ref{fig:gx_13+1_sequence} is therefore our best guess but only one of a few possible ways of incorporating them into the sequence. What does seem clear from inspection of the entire \gx\ data set is that these segments cannot plausibly be combined with other track segments---or accommodated in our sequence---as NB segments similar to the other ones we see. (We discuss these tracks further in Section~\ref{sec:discussion}).

The tracks in the HID look quite different from those of the other three sources. As we discuss below, different track branches can be identified in the HIDs; however, the fact that these branches are in general very broad and rather irregularly shaped makes them less useful for judging in what cases observations can be appropriately combined to form CD/HID tracks without significant secular shifts. The SID---although also displaying broad and somewhat irregular track branches---turned out to be more useful in this respect. To better illustrate the behavior of the source in the HID we show in Figure~\ref{fig:gx_13+1_colcode_example} a color-coded version of the track in panel C. The locations of the NB/FB vertex in the HID are in general less well defined than in the CD and they do not seem to follow a straight line.

\subsubsection{Secular Evolution}\label{sec:gx_13+1_evolution}

\begin{figure*}[t]
\centering
\includegraphics[height=9.3cm]{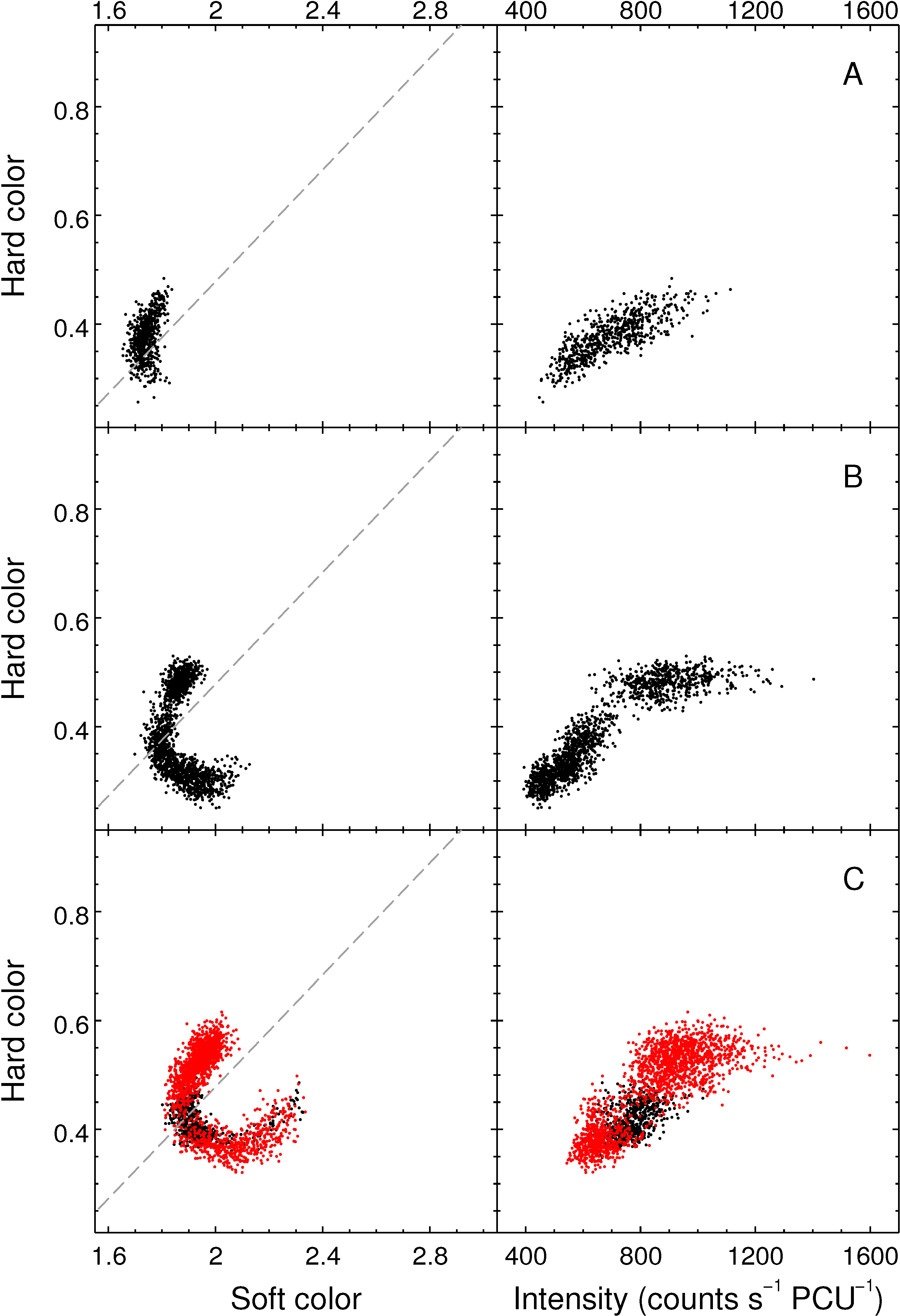}
\hspace{-0.035cm}
\includegraphics[height=9.3cm]{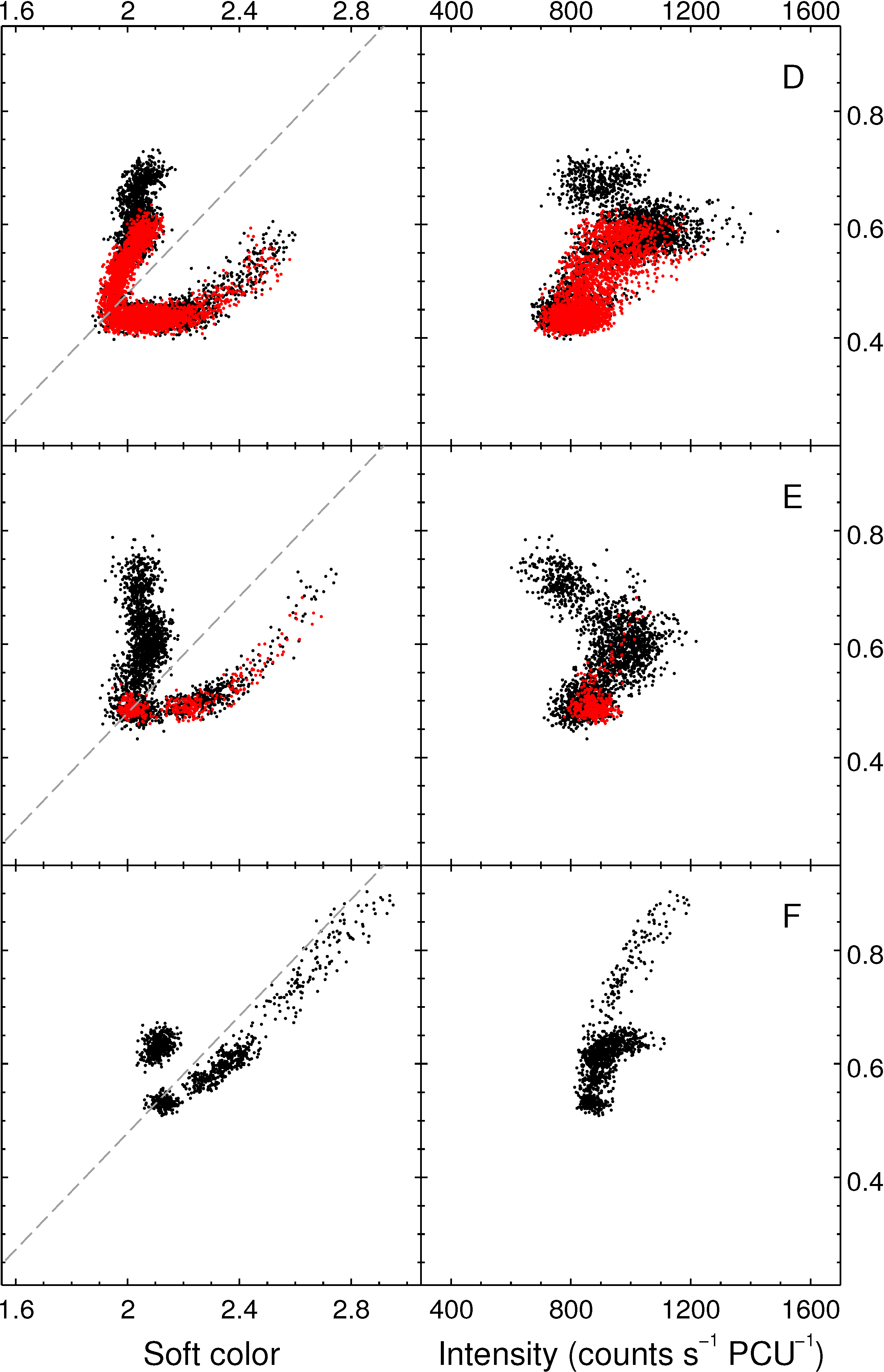}
\caption{A sequence of CDs/HIDs for \gx\ illustrating the secular evolution of the source. Table~\ref{tab:gx_13+1_obs} indicates the data used in each panel. In panels C, D, and E data from observations widely separated in time were combined; in each of these three panels a subset of the data---obtained over a period of at most $\sim$1.8~days---\,is shown in red. The dashed line in the CDs shows the approximate path followed by the NB/FB vertex.}\label{fig:gx_13+1_sequence}
\end{figure*}

\begin{deluxetable*}{clclc}[]
\tablewidth{18.0cm}
\tablecolumns{5}
\tablecaption{Time Intervals and Observations Used to Create \gx\ Tracks \label{tab:gx_13+1_obs}}
\tablehead{ & & \colhead{Interval Length} & & \colhead{Exp.\ Time} \\[0.3ex]
\colhead{\hspace{-0.13cm}Panel} & \colhead{\hspace{-0.65cm}Time Interval (MJD)} & \colhead{days (hr)\tablenotemark{a}} & \colhead{\hspace{-3.9cm}ObsIDs} & \colhead{ks (hr)\tablenotemark{b}}}
\startdata
A				& 51278.745--51278.971		& 0.23 (5.4)\phn\phn		& 40023-03-02-03						& 10.3 (2.9)\phn \bigstrut[b] \\
\hline
B				& 51276.746--51278.117		& 1.37 (32.9)\phn		& 40023-03-[01-00,01-02,02-(00:02),02-04]	& 26.0 (7.2)\phn \bigstrut \\
\hline
C				& 50990.407--50990.711		& 0.30 (7.3)\phn\phn		& 30051-01-09-[00:01]					& 9.8 (2.7) \bigstrut[t] \\[0.6ex]
				& 53767.006--53767.601	(R)	& 0.60 (14.3)\phn		& 91007-08-02-[00,000]					& 29.5 (8.2)\phn \bigstrut[b] \\
\hline
D				& 50981.074--50982.844	(R)	& 1.77 (42.5)\phn		& 30050-01-01-[04:08,050,080]				& 54.4 (15.1) \bigstrut[t] \\[0.6ex]
				& 50984.407--50984.576		& 0.17 (4.1)\phn\phn		& 30050-01-02-03						& 8.1 (2.3) \\[0.6ex]
				& 51390.995--51394.488		& 3.49 (83.8)\phn		& 40022-01-[01-(00:01),02-00,02-000]		& 39.0 (10.8) \\[0.6ex]
				& 51447.322--51447.428		& 0.11 (2.5)\phn\phn		& 40023-03-04-00						& 6.8 (1.9) \bigstrut[b] \\
\hline
E				& 51007.409--51007.560		& 0.15 (3.6)\phn\phn		& 30051-01-12-01						& 8.0 (2.2) \bigstrut[t] \\[0.6ex]
				& 53409.846--53409.949	(R)	& 0.10 (2.5)\phn\phn		& 90173-01-01-00						& 6.1 (1.7) \\[0.6ex]
				& 54740.627--54743.339		& 2.71 (65.1)\phn		& 93046-08-[02-00,03-(00:01)]				& 7.8 (2.2) \\[0.6ex]
				& 55409.062--55413.805		& 4.74 (113.8)			& 95338-01-[01-(00:07),03-00,03-05]			& 24.3 (6.8)\phn \bigstrut[b] \\
\hline
F				& 50997.607--51003.646		& 6.04 (144.9)			& 30051-01-[11-(00:03),12-00]				& 17.5 (4.8)\phn \bigstrut
\enddata
\tablenotetext{a}{The interval length is shown in units of both days and hours.}
\tablenotetext{b}{The total exposure time is shown in units of both ks and hours.}
\tablecomments{Subsets colored in red in Figure~\ref{fig:gx_13+1_sequence} are denoted by (R) in the Time Interval column. In the ObsIDs column a colon denotes a range.}
\end{deluxetable*}

\gx\ shows behavior that is in some ways quite different from that of the other three sources, especially in the HID. A more detailed description of the secular evolution of this source is therefore warranted. We note at the outset that, as Figure~\ref{fig:gx_13+1_all_data} indicates, the largest portion of the total exposure time for the source was spent on tracks near the middle of the sequence in Figure~\ref{fig:gx_13+1_sequence} (especially panel D), and it is therefore natural that the tracks in these panels would be the most complete ones. Conversely, the amount of data available at the lowest (panels A and B) and highest (panel F) vertex locations (all of which is shown in Figure~\ref{fig:gx_13+1_sequence}) is very small. These tracks therefore seem most likely to suffer from incompleteness.

We first focus on the evolution of the tracks in the CD. As the tracks move up the vertex line the FB gradually rotates counterclockwise and becomes longer and straighter (more Sco-like). Starting at the other end of the sequence, the NB gradually becomes longer as the tracks move down the vertex line from panel F to D. The NB seems to shorten again going from panel C to A, although we suspect that this may be due to incompleteness in panels A, B, and possibly C. These three tracks may also be missing HB and HB upturn portions.

\begin{figure}[t]
\centerline{\includegraphics[width=8.6cm]{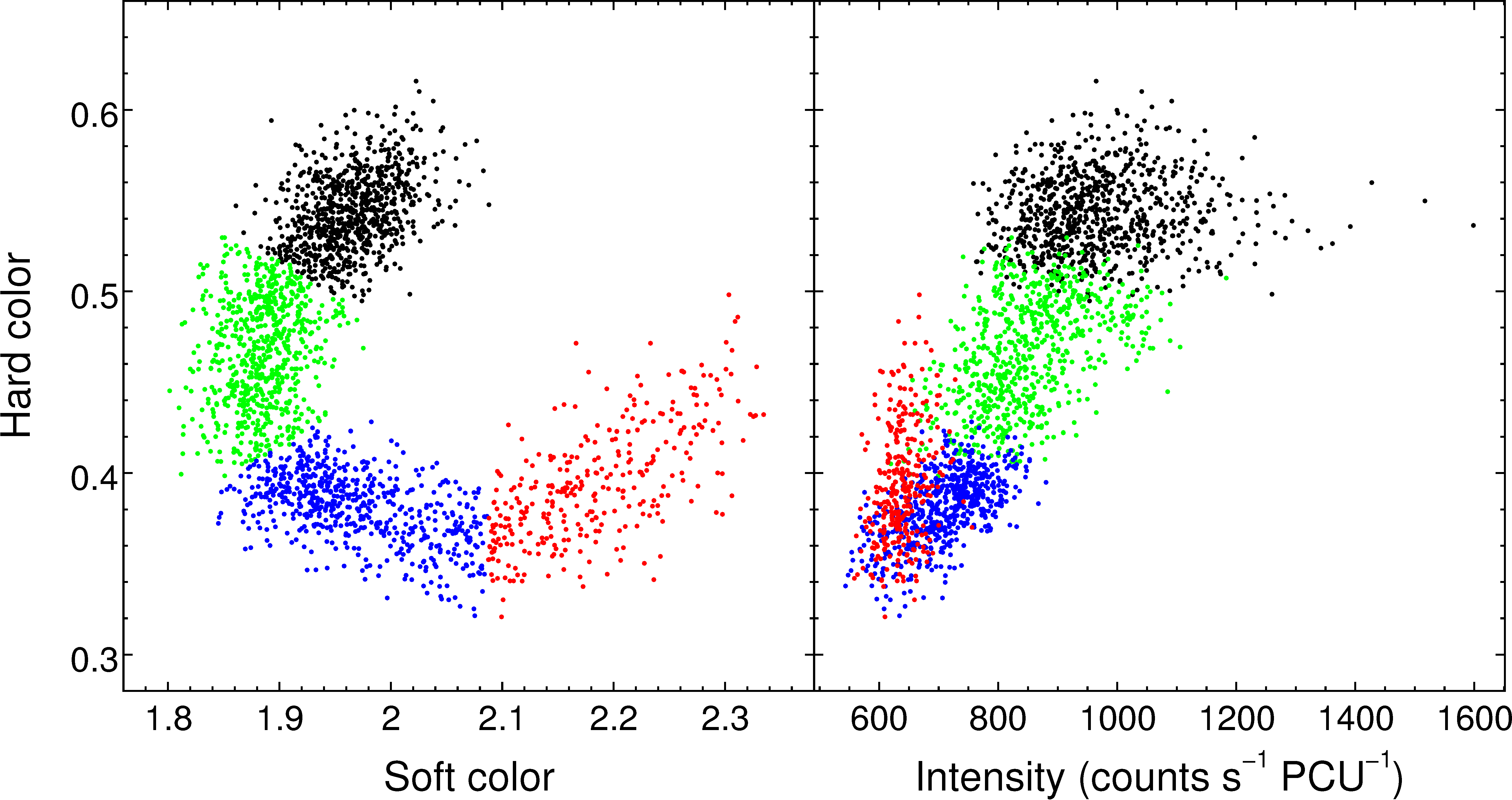}}
\caption{Color-coded version of the track in panel C in Figure~\ref{fig:gx_13+1_sequence}, illustrating the portions of the HID track corresponding to several segments along the track in the CD.}\label{fig:gx_13+1_colcode_example}
\end{figure}

The evolution in the HID is less obvious. At the lower vertex locations the NB is strongly tilted to the right and for a given hard-color value spans a large intensity range, especially in its upper part. The NB gradually becomes more compact going to higher vertex locations (although it is always broad compared to the NBs observed in the other three sources). The behavior of the FB in the HID is even harder to discern than that of the NB. The morphology of the FB in panel C can more easily be seen in the color-coded version in Figure~\ref{fig:gx_13+1_colcode_example}. The upward-pointing part of the FB is essentially vertical (but very short) in panel C, and is at the lowest intensities. Going counterclockwise along this track in the CD, the highest intensities are near the top of the NB and then the intensity becomes gradually lower moving along the track until the FB reaches its lowest point (in hard color), after which the intensity becomes approximately constant. (The same behavior is seen in panel B.) In panels D and E, the flat part of the FB in the CD forms a broad patch at the bottom of the track in the HID. The rising part of the FB extends from the left side of this patch, tilted slightly to the right from vertical, and overlaps with the broad NB (see, e.g., the red-colored data in panel E).

\subsubsection{Comparison with \src, \cyg, and \cir}\label{sec:comparison_gx}

The tracks in the CD of \gx\ show many similarities to those of \src, \cyg, and \cir. The evolution of the FB in the CD is similar to that seen for \src\ in selections A--H---with the FB rotating counterclockwise and growing longer (and more Sco-like)---and likewise similar to that of \cir. There are strong similarities between the shape of the FB of \gx\ in panels D--F and, e.g., the FB of \src\ in selections E--H, \cyg\ in panel J, and \cir\ in panels H and I. As in the other three sources, a lower vertex in the CD of \gx\ moves along a straight line with positive slope. The NB is also similar to the ones of the other three sources (being mostly straight and oriented up and to the right), and the possible HB and HB upturn segments in \gx\ have a similar appearance to those of the other sources as well.

Although less obvious at first sight, there are also similarities in the HID between GX~13+1 and the other three sources. The NB is oriented up and to the right, becoming gradually shorter and closer to vertical as the tracks move up in hard color---i.e., the intensity swings along the NB become smaller---and at the same time the maximum intensity observed on the NB gradually decreases (although for this to extend to panels A and B we need to assume they have incomplete NBs). This is also the case for the other three sources. What we interpret as traversals onto the HB and HB upturn in \gx\ are manifested as movement toward lower intensities and higher hard color; this is in general the case for Z sources, including the other three discussed in this paper. In the earlier panels in the \gx\ sequence the initial part of the FB is toward lower intensities (i.e., dipping FB behavior), as is seen in the other three sources. The \gx\ FB rotates slightly clockwise in the HID (similar to \cyg) as the source moves up the vertex line in the CD, evolving into an FB that shows an intensity increase rather than decrease. A notable difference between \gx\ and the other three sources is the movement of the lower vertex in the HID. The location of the lower vertex in the HID of \gx\ is in general rather poorly defined, and does not seem to follow a straight line. However, it seems clear that the lower vertices in panels A and B in Figure~\ref{fig:gx_13+1_sequence} are at a significantly lower intensity (\mbox{$\sim$550--600 counts s$^{-1}$ PCU$^{-1}$}) than the ones in panels C--F (\mbox{$\sim$800--900 counts s$^{-1}$ PCU$^{-1}$}), in contrast to the other three sources, where in general the intensity at the lower vertex decreases as it moves up the vertex line in the CD.

As is clear from the above \gx\ shows similarities to both Cyg-like and Sco-like Z sources. Overall, the earlier panels in the track sequence are more Cyg-like and the later ones more Sco-like, although the distinction between the two is less clear than for the other three sources.

\section{Discussion}\label{sec:discussion}

\subsection{Secular Evolution}\label{sec:secular_evolution}

The main goal of this paper is to study secular evolution in the CDs and HIDs of NS-LMXBs, using three sources historically known to show substantial changes in the shape and location of their CD/HID tracks:  \cyg, \cir, and \gx. In particular, we want to determine to what extent the secular evolution of these three sources is similar to that seen in \src, the first source found to evolve through all NS-LMXB subclasses.

\begin{figure*}[]
\vspace{0.1cm}
\centerline{
\hspace{0.915cm}
\includegraphics[width=13.8cm]{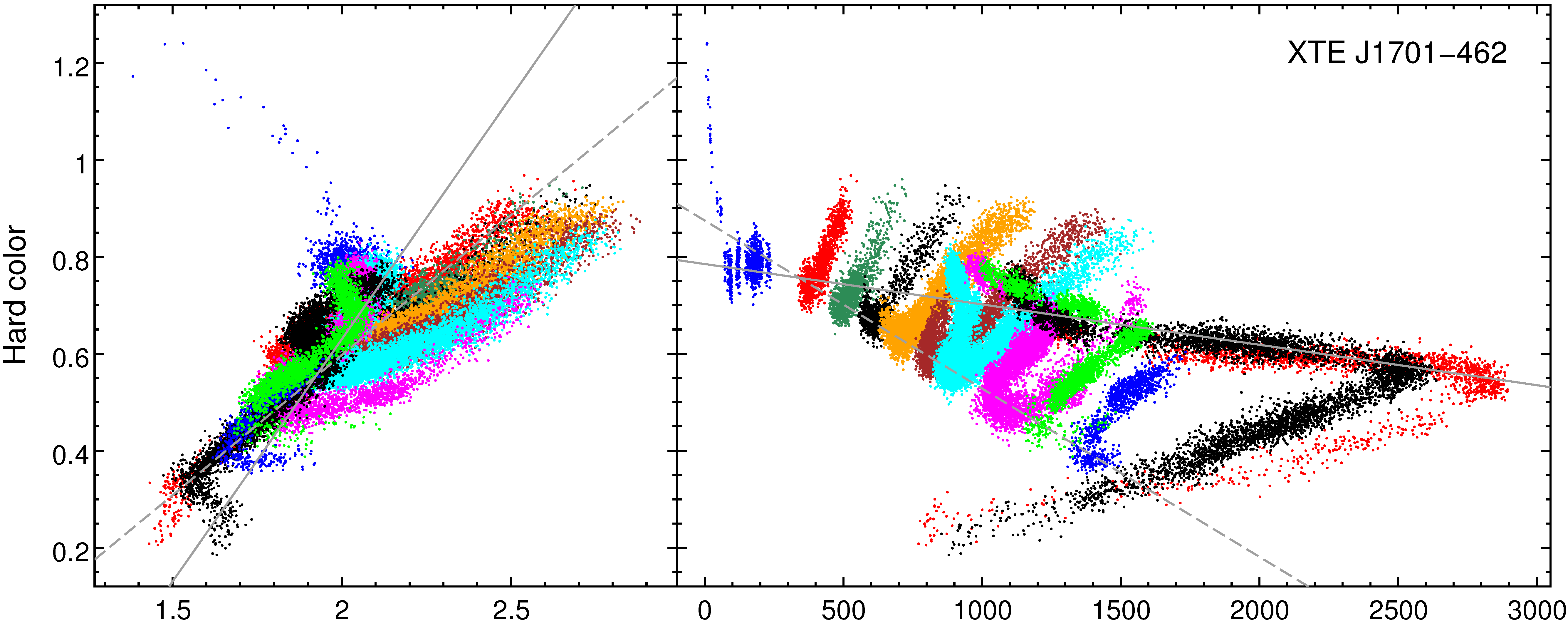}\\
}
\vspace{0.2cm}
\centerline{
\hspace{0.915cm}
\includegraphics[width=13.8cm]{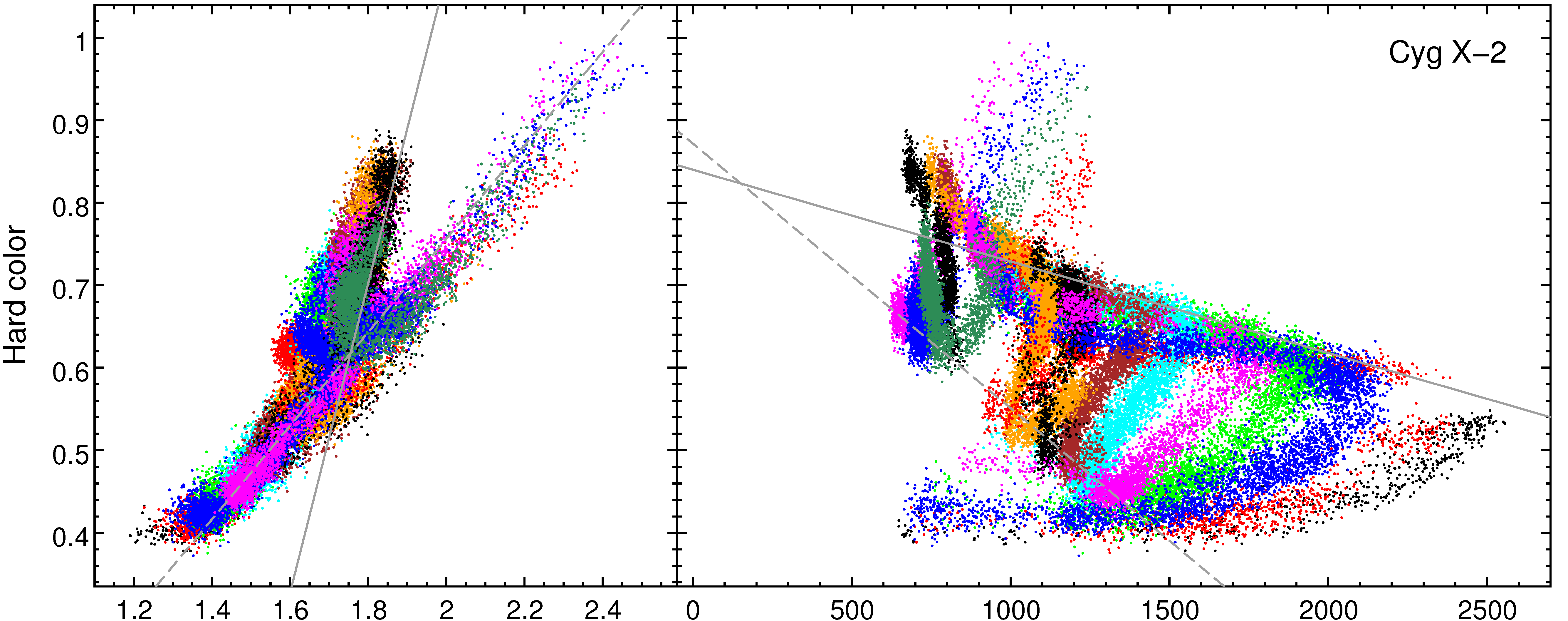}\\
}
\vspace{0.2cm}
\centerline{
\hspace{0.915cm}
\includegraphics[width=13.8cm]{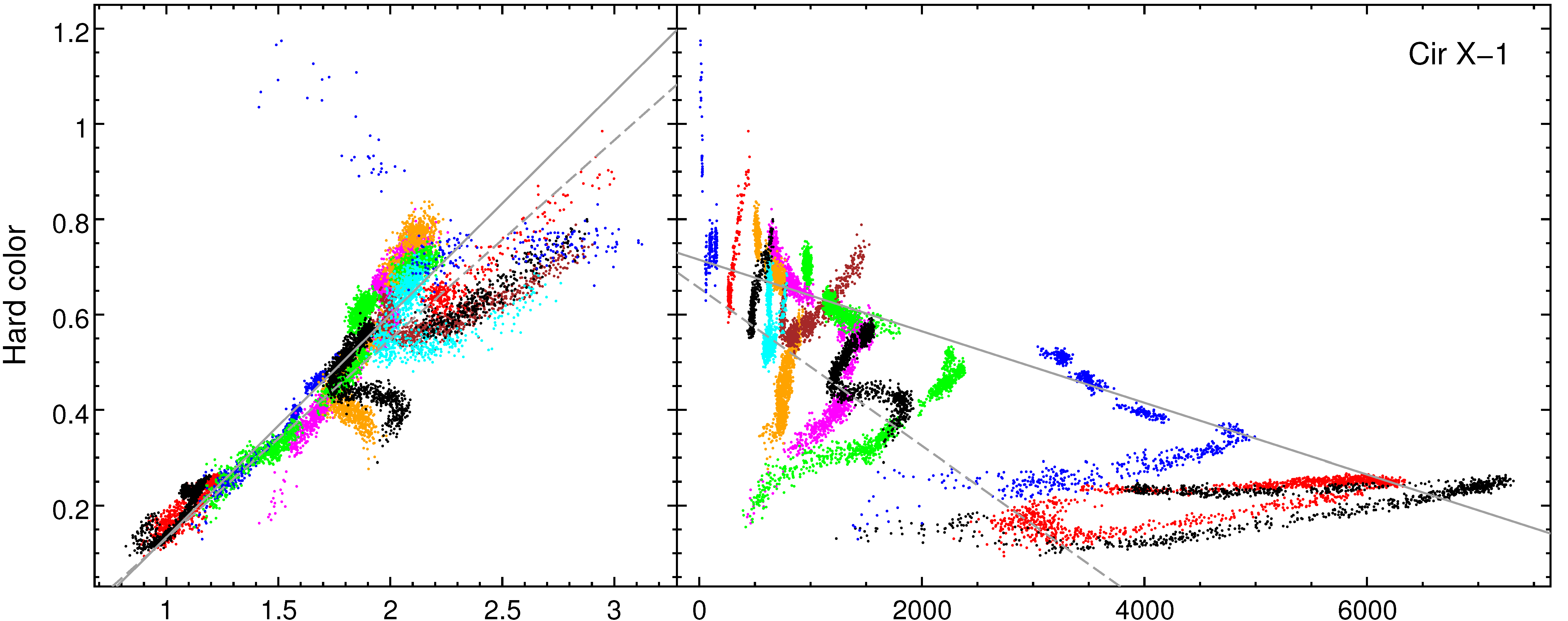}\\
}
\vspace{0.2cm}
\centerline{
\hspace{0.915cm}
\includegraphics[width=13.8cm]{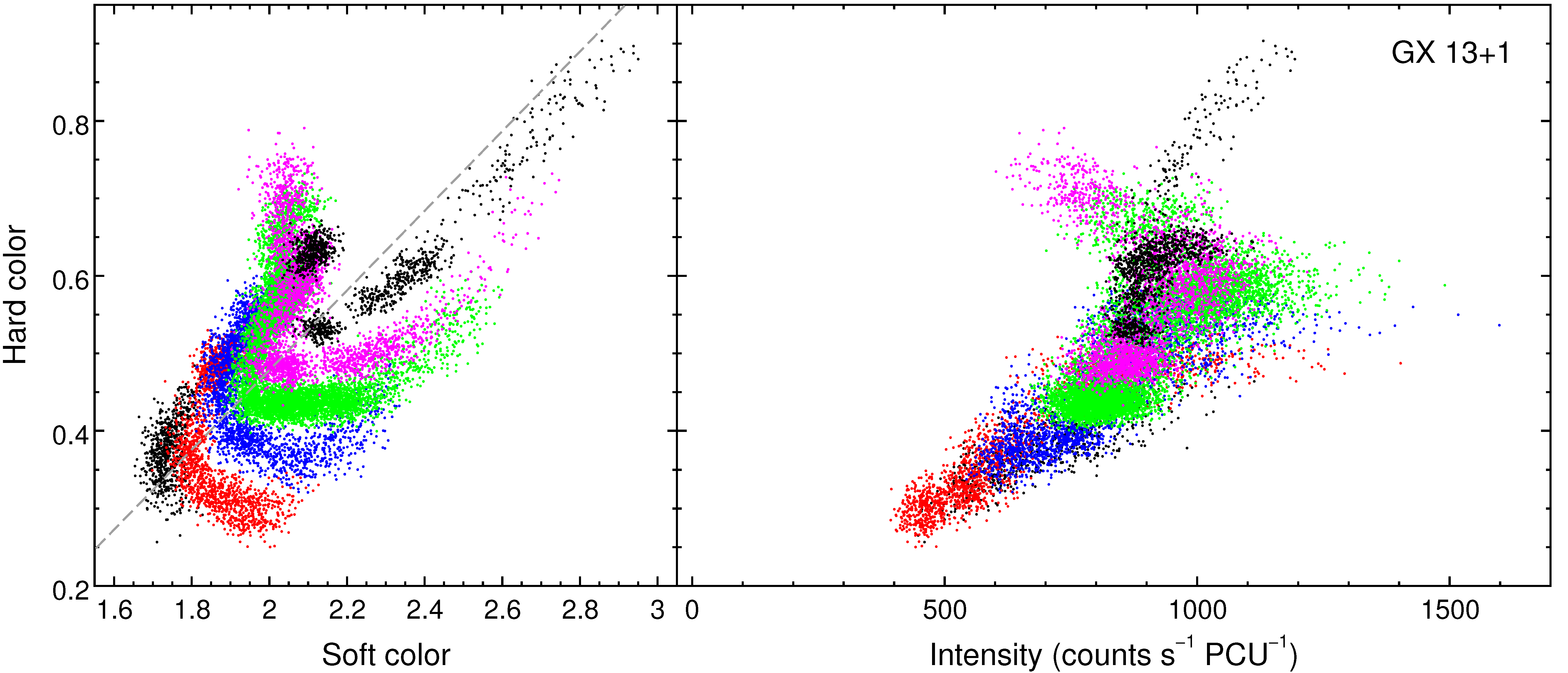}\\
}
\caption{CDs/HIDs illustrating the overall secular evolution of \src, \cyg, \cir, and \gx. In the panels we show (from top to bottom) all the tracks from Figures~\ref{fig:1701_sequence}, \ref{fig:cyg_x-2_sequence}, \ref{fig:cir_x-1_sequence}, and \ref{fig:gx_13+1_sequence}, respectively. Different colors are used to distinguish between the individual tracks in each panel.}\label{fig:all_tracks_single_plot}
\end{figure*}

In Figure~\ref{fig:all_tracks_single_plot} we provide an overview of the secular evolution of the CD/HID tracks of \src, \cyg, \cir, and \gx. In each of the four panels we show all tracks from the sequences in Figures~\ref{fig:1701_sequence}, \ref{fig:cyg_x-2_sequence}, \ref{fig:cir_x-1_sequence}, and \ref{fig:gx_13+1_sequence} in a single CD/HID. (For \src\ we show only the red-colored subsets from Figure~\ref{fig:1701_sequence}, where applicable.) Strong secular evolution is found in all four sources, consistent with reports in the literature. While for \src\ all data come from the dense monitoring of a single 19~month outburst, the data for the other three sources were collected sporadically over a time span of $\sim$14--15~years. As a result, the secular evolution in these sources could not be followed in ``real time," as was possible for \src, but had to be reconstructed from multiple isolated (partial) tracks. To some extent this limited our ability to compare the secular evolution between sources, in particular for \gx. Despite these issues, it is obvious from Figure~\ref{fig:all_tracks_single_plot} that the secular evolution of the four sources has many common characteristics.

In the following we summarize our key findings, focusing on the similarities between the systems.

\begin{enumerate}

\item As part of their secular evolution, we see clear and continuous transitions between different NS-LMXB subclasses in all four sources. While some of the behavior we report has been described in previous works, our work for the first time unambigously links strong secular evolution to transitions between various NS-LMXB subclasses in \cyg, \cir, and \gx.  Of the four sources \src\ and \cir\ have shown the largest range in behavior; they have displayed Cyg-like and Sco-like Z tracks, atoll soft and hard states, and have both at some point entered quiescence or near-quiescence (\mbox{$\sim$$10^{35}$ erg s$^{-1}$} in the case of \cir; \citealt{sell2010}).  \cyg\ and \gx\ have only shown Cyg-like and Sco-like Z tracks, with the secular changes in \gx\ being the most moderate of the four sources. 

\item{Cyg-like Z source behavior (with large variations in intensity along the NB and HB, and a ``dipping'' FB that is directed toward lower intensities) is observed at the highest overall intensities, Sco-like Z source behavior (with small intensity variations along the NB and, if present, the HB, and large increases in intensity along the FB) at lower intensities, and (in \src\ and \cir) atoll behavior at the lowest overall intensities.}

\item{Although possible incompleteness in some of the CD/HID tracks clouds the picture somewhat, the order in which the Z source branches evolve and disappear in \cyg\ and \cir\ seems to be consistent with what is seen in \src. As the overall intensity decreases, the HB shortens and disappears first, followed by the NB, and finally the FB. The situation is less clear for \gx; the earlier tracks in Figure~\ref{fig:gx_13+1_sequence} would need to be missing an HB (and possibly part of the NB) for the behavior of \gx\ to be consistent with \src\ as well.}

\item{As the shapes of the Z tracks and branches gradually change, the vertices of the tracks shift along well-defined lines in the CD, moving toward higher color values as the overall intensity decreases. (We note, however, that the overall intensity evolution is less clear-cut for \gx, and that \cir\ shows some deviations from the general intensity trend.) For \src\ and \cyg\ this also applies to the HID, and to a large extent for \cir\ as well.}

\item{As the overall intensity of the Z tracks increases and they become spectrally softer, their dynamic range (i.e., ratio of maximum to minimum intensity) increases as well. Whether this trend also extends to the tracks at the lowest color values in \gx\ is unclear due to possible incompleteness issues.}

\end{enumerate}

Despite the fact that differences exist in their detailed behavior, based on the above findings we conclude that the secular evolution in the four NS-LMXBs that we studied largely follows similar patterns. This strongly suggests that the sequences of NS-LMXB subclasses reported in this work are representative of the class of NS-LMXB as a whole.

\subsection{Evolution of the Flaring Branch}\label{sec:flaring_branch_evolution}

Of the three main Z source branches, the FB undergoes the most dramatic changes in its shape and orientation. In all four sources it rotates in both the CD and HID and evolves from a branch that shows a strong decrease in intensity with respect to the NB/FB vertex (in the Cyg-like Z tracks), to one that shows a strong increase in intensity with respect to that vertex (in the Sco-like Z tracks), although the amplitude of these intensity changes is significantly smaller in \gx\ than in the other three sources.  From the dense ``coverage'' of the secular evolution of \src\ and \cyg\ it is clear that this FB evolution from dipping to flaring is a very gradual process, with the morphology changing smoothly throughout the track sequences.  This suggests that the dipping Cyg-like FB and the Sco-like FB are related phenomena, and it seems unlikely that the overall observed FB behavior can be explained by two presumably unconnected mechanisms: absorption by the outer disk, which was proposed as an explanation for the \cyg\ dipping behavior by \citet{balucinska-church2011}, and nuclear burning on the neutron star (combined with increases in the mass accretion rate in the Sco-like Z sources), proposed as an explanation for the flaring FB by \citet{church2012}. (See also \citealt{balucinska-church2012}.)

\subsection{Other Z Source Transients}\label{sec:other_z_transients}

We note that two additional Z source transients discovered after we began work on this paper, \mbox{IGR J17480--2446} (also known as \mbox{Terzan 5 X-2}) and \maxi, have shown similar behavior to the sources studied here. \xte\ observations of \mbox{IGR J17480--2446} revealed clear Cyg-like Z source behavior at the peak of its $\sim$2.5~month outburst and atoll behavior at lower intensities; due to the rapid source evolution and sparsity of the data Sco-like Z behavior could not be identified unambiguously (\citealt{altamirano2010}; \citealt{chakraborty2011}; \citealt{altamirano2012}; D.~Altamirano et al.\ 2015, in preparation).

\maxi\ showed both Cyg-like and Sco-like Z behavior during its 16~month outburst (\citealt{homan2011}; \citealt{sugizaki2013}; \citealt{homan2014}; J.~Homan et al.\ 2015, in preparation). Figure~2 in \citet{homan2014} shows a combined HID using all the \xte\ data for the source (panel a) as well as four time-selected HID tracks illustrating the secular evolution of the source (panels b--e). The total HID looks similar to that of \cyg\ (Figure~\ref{fig:cyg_x-2_all_data}). The higher-intensity \maxi\ tracks (panels b and c) are Cyg-like Z tracks; the track in panel b shows an extended dipping FB and resembles the highest-intensity tracks of \src, \cyg, and \cir. The lower-intensity tracks of \maxi\ (panels d and e) are Sco-like Z tracks, resembling the tracks in panels F and G--H, respectively, in the \src\ sequence, and \mbox{K+L} and M in the \cyg\ sequence. As the overall intensity changes and the \maxi\ tracks gradually evolve in shape, both the upper and lower vertices follow straight lines in the HID, as observed for \src\ and \cyg. The \xte\ coverage of \maxi\ ended before the outburst came to an end, while the source was still showing Sco-like Z behavior. It may subsequently have passed through the atoll soft and hard states on its way back to quiescence.
 
Finally, we note that the atoll transient \mbox{XTE J1806--246} likely entered the Sco-like Z source regime briefly at the peak of its outburst in 1998 \citep{wijnands1999}. Observations during the outburst peak show the source tracing out a curved (partial) track resembling the lower NB+FB part of a Sco-like Z track. As one would expect for a transition from the atoll to Z regime, this track is at lower color values (and higher intensities) with respect to the observations in the atoll soft state taking place before and after the outburst peak. While the source traced out this track, a \mbox{7--14 Hz} QPO was detected, whose frequency increases as a function of position along the track (going from the NB to FB), consistent with the behavior of the normal/flaring-branch oscillation in Z sources.

The three sources discussed in this section, along with the four sources analyzed in this paper, constitute the entire sample of NS-LMXBs known or believed to have shown transitions between at least two of the three main NS-LMXB subclasses (Cyg-like Z, Sco-like Z, atoll). The fact that these three additional transients show behavior seemingly largely consistent with the other four sources (e.g., the relative luminosity dependence of the subclasses) strengthens our conclusion at the end of Section~\ref{sec:secular_evolution}.

 \subsection{Role of the Mass Accretion Rate}\label{sec:mass_accretion_rate}
 
The spectral fitting results of \citet{lin2009a} indicate that the secular evolution of \src\ was driven by changes in the mass accretion rate---they find that the $\dot{M}$ inferred from the accretion disk component in the spectra at the upper and lower vertices decreases monotonically when moving left along the two vertex lines in the HID. (See also discussion in \citetalias{homan2010}.) Based on the similarities in secular behavior between \src\ and the three sources studied in this paper we conjecture that the secular evolution of \cyg, \cir, and \gx\ is also driven by changes in the mass accretion rate. The case is clearest for \cyg, whose behavior is most regular of the three sources and closest to that of \src. The fact that the overall intensity of the \cyg\ tracks decreases monotonically as they gradually evolve in shape and move along the vertex lines to higher color values makes it plausible that the mass accretion rate in \cyg\ decreases monotonically along the sequence, as seems to be the case for \src.

Although there are some important differences between the evolution of \cir\ and \src, the global properties are still very similar.  The overall large intensity decrease from the Cyg-like tracks in the earlier panels to the Sco-like (and finally atoll) tracks in the later panels suggests a decreasing mass accretion rate going from panel A to L. However, the weaker correlation between track shape and location in the HID compared to the other sources---in particular manifested in the jumping back and forth in the HID seen during part of the sequence (panels F--I)---suggests that the mass accretion rate progression may not be strictly monotonic along the sequence. In Section~\ref{sec:cir_x-1_discussion} we suggest that these ``discrepancies'' between the track shape and location in the HID may be the result of rapid changes in the mass accretion rate.

Despite the differences in the behavior of \gx\ compared to the other three sources we find it most likely that changes in the mass accretion rate are also responsible for the secular evolution of \gx\ and that, as in the other sources, $\dot{M}$ decreases as the source moves up the vertex line in the CD---i.e., going from panel A to F in Figure~\ref{fig:gx_13+1_sequence}. As the tracks become overall spectrally harder the maximum intensity of a given track decreases, and the intensity swings along the track become smaller, as observed for the other three sources, although incompleteness in the tracks in panels A and B needs to be invoked to explain their deviations from this trend. Such incompleteness can, however, not explain the fact that the lower vertex in the HID is at lower intensities in those two tracks than in tracks that we suggest exhibit lower mass accretion rates, opposite to the behavior of \src.

Our results support the idea that the mass accretion rate is responsible for changes between subclasses in individual sources, with Sco-like Z behavior appearing at lower accretion rates and overall luminosities than Cyg-like Z behavior (and atoll behavior appearing at accretion rates that are lower still). However, in this picture other parameters, such as the neutron star spin, mass, and magnetic field, may still affect the luminosity scale of the subclass sequence---i.e., the range in luminosity (and mass accretion rate) over which a given subclass appears may differ between sources.

\citet{lin2009b,lin2009a} showed that for its best-estimate distance of \mbox{8.8 kpc}, \src\ reached super-Eddington luminosities during its Cyg-like stage. Even for a distance that is 30\% smaller this would have been true. The super-Eddington nature of the Cyg-like tracks is further supported by observations of two Sco-like Z sources with accurately known distances, \mbox{Sco X-1} (\mbox{2.8$\pm$0.3 kpc}; \citealt{bradshaw1999}) and \mbox{LMC X-2} (\mbox{50$\pm$2 kpc}; \citealt{pietrzynski2013}). Spectral fits indicate that those sources are either around Eddington or mildly super-Eddington \citep{bradshaw2003,agrawal2009}. Given that Cyg-like behavior is observed at significantly higher overall intensities than Sco-like behavior in \src, \cyg, and \cir, this suggests that Cyg-like Z tracks may in general exhibit luminosities well above the Eddington limit.

As the sources analyzed in this paper evolve from Sco-like to Cyg-like Z behavior, it is interesting to see a large increase in the dynamic range (i.e., ratio of maximum to minimum intensity) of the Z tracks. This is most clearly seen in the tracks of \src, \cyg, and \cir, although this also holds for \gx, apart from the possibly very incomplete track in panel A (Figure~\ref{fig:gx_13+1_sequence}). The most extreme intensity swings are seen in \cir, where we observe changes in the intensity (going between the tip of the dipping FB and the upper vertex) by factors up to 8 in periods as short as 20~minutes in the highest-intensity tracks. In these extreme Cyg-like Z tracks, \cir\ seems likely to have reached significantly higher luminosities (perhaps by a factor of $\sim$3) and exhibited significantly higher mass accretion rates than observed in \src\ at the peak of its outburst. Given that the large Cyg-like intensity swings observed in \cir\ (as well as \src\ and \cyg) occur in inferred luminosity ranges that are near- or super-Eddington, it seems likely that radiation pressure effects play an important role. While it is unlikely that the mass accretion rate toward the inner parts of the accretion flow varies this strongly on such a short timescale, strong radiation pressure may result in substantial changes in the geometry of the inner accretion flow, with a (large) fraction of the mass inflow possibly being redirected into wind or jet outflows, resulting in large intensity swings.

Given the near-/super-Eddington nature of the Z sources, it is also interesting to compare them with the transient ultraluminous X-ray source \mbox{M82 X-2}. This source was recently identified as an accreting neutron star \citep{bachetti2014} and has been seen to reach a maximum luminosity of \mbox{$\sim$1.8$\times10^{40}$~erg~s$^{-1}$} \citep{feng2010}, which is a factor of  $\sim$30 higher than the peak luminosity of \src\ \citep{lin2009a}. However, the short-term light curves of \mbox{M82 X-2} presented in \citet{kong2007} do not show any of the strong luminosity swings that characterize the Cyg-like Z sources. A possible explanation for this is that due to the much higher neutron star magnetic field in \mbox{M82 X-2} (as inferred from the strong pulsations in the light curve; \citealt{bachetti2014}) the accretion flow geometry in that source is very different from that in the Cyg-like Z sources, perhaps preventing the strong feedback mechanisms that we think may cause the large luminosity swings.

Three of the sources studied in this paper (\src, \cyg, and \gx) were also part of a large-sample study of NS-LMXB behavior by \citet{munoz2014}. These authors find that the Cyg-like and Sco-like Z sources occupy a region of the rms--luminosity diagram that partly overlaps with that of the luminous black hole X-ray binary \mbox{GRS 1915+105}. We note, however, that overall the Z sources and \mbox{GRS 1915+105} behave very differently \citep[e.g.,][]{belloni2000}, suggesting that at near/super-Eddington luminosities the observed properties of NS-LMXBs and black hole X-ray binaries are quite different, perhaps (in part) due to the increased importance of radiation feedback from the neutron star surface.

\subsection{Nonstationary Accretion in \cir}\label{sec:cir_x-1_discussion}

As discussed in Sections~\ref{sec:cir_x-1} and \ref{sec:mass_accretion_rate}, the secular behavior of \cir\ is not as regular as that of \src\ and \cyg, as for example manifested in deviations of the track vertices from a straight line path (especially in the HID). We suspect that this may be the result of the highly nonstationary accretion in \cir, thought to arise from the binary's eccentric orbit. From transient LMXBs it is known that highly nonstationary accretion can lead to substantial hysteresis (e.g., \citealt{maccarone2003,homan2005}), with certain source states being observed over a much larger luminosity range than would be the case for slowly changing accretion rates. \citet{yu2009} showed that these effects become stronger as the changes in mass accretion rate become faster. We speculate that, analogously, highly nonstationary accretion could be a possible explanation for certain Z track shapes being observed over a range of luminosities, rather than at a single location along the vertex lines. While all four sources in our sample undergo nonstationary accretion, which gives rise to their secular evolution, the rate at which the mass accretion rate varies with time is likely considerably higher in \cir\ than in the other three sources. This is suggested by the fastest timescales on which we observe secular evolution  in \cir, which are shorter than in the other sources. The fastest secular evolution seen for \src\ took place during the earliest stages of the outburst (panels A--C). Inspection of the SID indicates that the lower vertex of the HID track in panel A is at \mbox{$\sim$1900 counts s$^{-1}$ PCU$^{-1}$} rather than the \mbox{$\sim$1600 counts s$^{-1}$ PCU$^{-1}$} implied by the vertex line shown; we see indications for a similar, but smaller, shift in panel B. We also note that \maxi\ showed strong deviations from a straight vertex line (which it otherwise followed) during the rapid rise of its outburst (J.~Homan et al.\ 2015, in preparation). This lends support to our hypothesis that rapid changes in the mass accretion rate underlie the observed deviations in \cir.

An additional factor that may play a role in the somewhat irregular behavior of \cir\ is the fact that the neutron star spin, the companion spin, and the orbital axis are likely mutually misaligned in this system \citep{brpo1995,heinz2013}. This can lead to precession of the neutron star spin axis, which may in turn affect the rotation of the inner accretion flow. Changes in the projected area of the inner accretion flow might then result in similar track shapes being observed at different X-ray fluxes.

\subsection{The Z Source Classification of \gx}\label{sec:gx_13+1_discussion}

Overall, we find that the shape and evolution of the CD/HID tracks of \gx\ show much stronger similarities to the other three sources analyzed in this paper than to atoll sources. Most generally, just the fact that we see gradual evolution in the shape of large-scale tracks traced out by the source as the locations of the tracks shift over a considerable range in the CD and HID points to a Z source, since such secular evolution in the CD/HID has never been observed in atoll sources---only in Z sources. Moreover, the specifics of this evolution show many similarities to those of \src, \cyg, and \cir, as discussed in Section~\ref{sec:gx_13+1}. Furthermore, our identification of track segments in the \gx\ sequence (panels D and E in Figure~\ref{fig:gx_13+1_sequence}) as the HB and HB upturn (see discussion in Section~\ref{sec:gx_13+1_analysis}) is supported by the detection of a highly significant \mbox{$\sim$22--29 Hz} QPO in the corresponding timing data of these segments (J.~K.~Fridriksson et al.\ 2015, in preparation). These frequencies are considerably lower than the \mbox{57--69 Hz} QPO found by \citet{homan1998} in parts of the tracks that we now classify as the upper part of the NB (see also \citealt{schnerr2003}). If we interpret the QPO as an HBO, the decrease in its frequency as the source moves onto the HB from the upper NB is consistent with what has been seen in other Z sources (see, e.g., \citealt{jonker2000} and \citealt{homan2002}).

\citet{munoz2014} find that in their source sample there are systematic differences between the ranges in fast-variability level (broadband fractional rms) shown by the different NS-LMXB subclasses. The Cyg-like Z sources reach the highest fractional rms values (up to $\sim$10\%--15\%), followed by the Sco-like Z sources (with peak values in the $\sim$5\%--10\% range), while bright persistent atoll sources and transient atolls in a bright soft state are typically observed at $\lesssim$5\%. \citet{munoz2014} find that the rms values oberved for \src\ in the different subclasses are consistent with this. The highest fractional rms in the Z sources is observed on the HB, and we find that on the presumed HB segments \gx\ reaches fractional rms values as high as $\sim$9\%--10\% (J.~K.~Fridriksson et al.\ 2015, in preparation). This seems more consistent with Z than bright atoll behavior. We also note that the hard color range over which these \gx\ HB segments are observed ($\sim$0.6--0.8) is consistent with HBs observed in both persistent and transient Z sources, but clearly too low to be consistent with the atoll hard state \citep{mythesis2011}.

Considering, in addition to the above, the timescale on which the \gx\ tracks are traced out (hours to $\sim$1~day instead of the days/weeks timescales usually observed in atoll sources), the extreme scarcity of observed type~I X-ray bursts from the source \citep{fleischman1985,matsuba1995,galloway2008}, and the reported radio brightness/behavior \citep{fender2000,homan2004} we confidently classify \gx\ as a Z source, one that switches between Sco- and Cyg-like Z behavior. However, the behavior of the source in the HID remains somewhat puzzling when compared with \src, \cyg, and \cir---especially the broad appearance of the tracks and the fact that in the spectrally harder tracks the NB/FB vertex is at higher, not lower, intensities than in the softer tracks.

This behavior may be connected to the fact that \gx\ is probably viewed at a high inclination angle (perhaps \mbox{$60^\circ$--\,$80^\circ$}) as indicated by observations of absorption dips in this system on several occasions \citep{diaz2012,dai2014,iaria2014}. It was recently shown for black hole X-ray transients that the shape of their HID tracks depends on the inclination angle \citep{munoz2013}. While it is not fully understood how inclination affects the shape of the HID, it is likely that an inclination dependence of the HID is present in NS-LMXBs as well, with different physical components, such as the accretion disk and boundary layer, being affected differently by changes in the viewing angle.

\section{Summary and Conclusions}\label{sec:summary}

It was recently shown that during its 2006--2007 outburst, the super-Eddington neutron star transient XTE J1701$-$462 evolved through all subclasses of low-magnetic-field NS-LMXBs as a result of changes in a single physical parameter: the mass accretion rate (\citealt{lin2009a}; \citetalias{homan2010}). The main goal of this paper is to study secular evolution in the CDs and HIDs of NS-LMXBs and, more specifically, to investigate whether evolution similar to that of \src\ occurs in other sources. To this end we studied the evolution of the CD/HID tracks of three sources that show complicated secular behavior---\cyg, \cir, and \gx---using all \xte\ PCA data obtained during the lifetime of the mission. No comprehensive study of the CD/HID behavior of these sources---using most or all of the \xte\ data now available---has previously been performed. We created sequences of CD/HID tracks---in many cases carefully piecing together data obtained throughout the \xte\ mission to construct more complete individual tracks than otherwise available---that demonstrate the secular progression in the track morphologies and locations. In the case of \cir, this analysis had to be preceded by the filtering out of data affected by intrinsic absorption, which otherwise strongly influences the appearance of the CD/HID.

We find that \cyg\ and \cir\ show strong similarities in their secular behavior to \src; in particular, both sources display gradual evolution between Cyg- and Sco-like Z tracks, with the latter occuring at lower intensities. \cir\ shows especially extreme versions of Cyg-like tracks at the highest observed intensities, with very large and rapid intensity swings but only small changes in spectral color. At the lowest observed intensities we also see clear transitions in the CD/HID from the atoll soft to hard state for \cir. The very gradual evolution of the FB morphology that is observed (especially in \src\ and \cyg), from a dipping (Cyg-like) FB to a flaring Sco-like one, suggests that these two different forms of the FB are related phenomena.

Although \gx\ shows behavior that is in some ways peculiar---especially in the HID---we find that the source also displays similarities to \src, \cyg, and \cir, and overall we conclude that the properties of \gx\ are strongly indicative of a Z source. 

The fact that \cyg\ and \cir\ show evolution between different NS-LMXB subclasses that is similar to the evolution seen in \src\ lends support to the suggestion of \citetalias{homan2010} that the behavior of \src\ is representative of the entire class of NS-LMXBs. It also strengthens their conclusion that (at least within individual sources) Cyg-like Z behavior takes place at higher luminosities and mass accretion rates than Sco-like Z behavior. Based on the similarities to \src\ we conjecture that the secular evolution of \cyg, \cir, and \gx\ is due to changes in the mass accretion rate in the systems, and overall our results support the notion that differences in mass accretion rate are the primary factor underlying the existence of the various NS-LMXB subclasses. We conclude that \cyg\ and \cir---like \src---probably reach super-Eddington luminosities, with \cir\ likely reaching significantly higher luminosities and mass accretion rates than the other sources.

\acknowledgements

We thank the referee for constructive comments. This research has made use of data obtained from the High Energy Astrophysics Science Archive Research Center (HEASARC), provided by NASA's Goddard Space Flight Center.

\bibliography{references}

\end{document}